\documentclass[%
 preprint,
 amsmath,amssymb,
 aps,
]{revtex4-2}

\usepackage{graphicx} 
\usepackage{dcolumn}  
\usepackage{bm}       
\usepackage{hyperref} 

\usepackage{caption}
\usepackage{subcaption}
\newsavebox{\largestimage}

\usepackage{physics}

\newcolumntype{C}[1]{>{\centering\arraybackslash}p{#1}}

\usepackage{xcolor}

\begin{document}

\preprint{}

\title{Injection locking at fractional frequencies\\of magnetic tunnel junction (MTJ)-based read sensors'\\ferromagnetic resonance modes}

\author{Ekaterina Auerbach}
 \email{auerbach@ifp.tuwien.ac.at}
 \affiliation{Institute of Solid-State Physics, Vienna University of Technology, Vienna 1040, Austria}

\author{Dmitry Berkov}
 \email{d.berkov@general-numerics-rl.de}
 \affiliation{General Numerics Research Lab e.V., Moritz-von-Rohr-Stra{\ss}e 1A, Jena 07745, Germany}

\author{Bernhard Pichler}
 \affiliation{Institute of Electrodynamics, Microwave and Circuit Engineering, Vienna University of Technology, Vienna 1040, Austria}

\author{Norbert Leder}
 \affiliation{Institute of Electrodynamics, Microwave and Circuit Engineering, Vienna University of Technology, Vienna 1040, Austria}
 
\author{Savas Gider}
 \affiliation{Western Digital Corporation, San Jose, CA 95119, USA}
 
\author{Holger Arthaber}
 \affiliation{Institute of Electrodynamics, Microwave and Circuit Engineering, Vienna University of Technology, Vienna 1040, Austria}

\date{\today} 

\begin{abstract}

Being nonlinear dynamic systems, magnetic read sensors should respond to an excitation signal of a frequency considerably different from their natural ferromagnetic resonance (FMR) frequencies. Because of the magnetization dynamics' inherent nonlinear nature, the sensors' response should be measured at the DC, excitation frequency, and its multiples (harmonics). In this paper, we present results of such measurements, accomplished using a one-port nonlinear vector network analyzer (NVNA), which show distinct resonances at fractional frequencies of the free layer (FL) FMR mode. Identification of these resonances, resulting from the nonlinear nature of the spin-torque (ST)-induced magnetization dynamics, was performed using micromagnetic modeling. In particular, we show that the measured DC response at the above-mentioned fractional frequencies can be explained by a low-order nonlinearity and strong magnetodipolar feedback between magnetic layers adjacent to an MgO barrier. Additionally, we determined that the simulated harmonic response is strongly enhanced by the mutual ST effect between these layers. Finally, we demonstrate that the read sensors' nonlinear magnetization dynamics and, by extension, their harmonic response are highly sensitive to various magnetic and ST parameters. Thus, this study shows that using NVNA measurements in conjunction with micromagnetic modeling can clarify the uncertainty in the definition of these parameters.

\end{abstract}

\maketitle


\section{\label{sec_introduction}Introduction}

\begin{figure}[!htb]
  \centering
    \includegraphics[scale=.07]{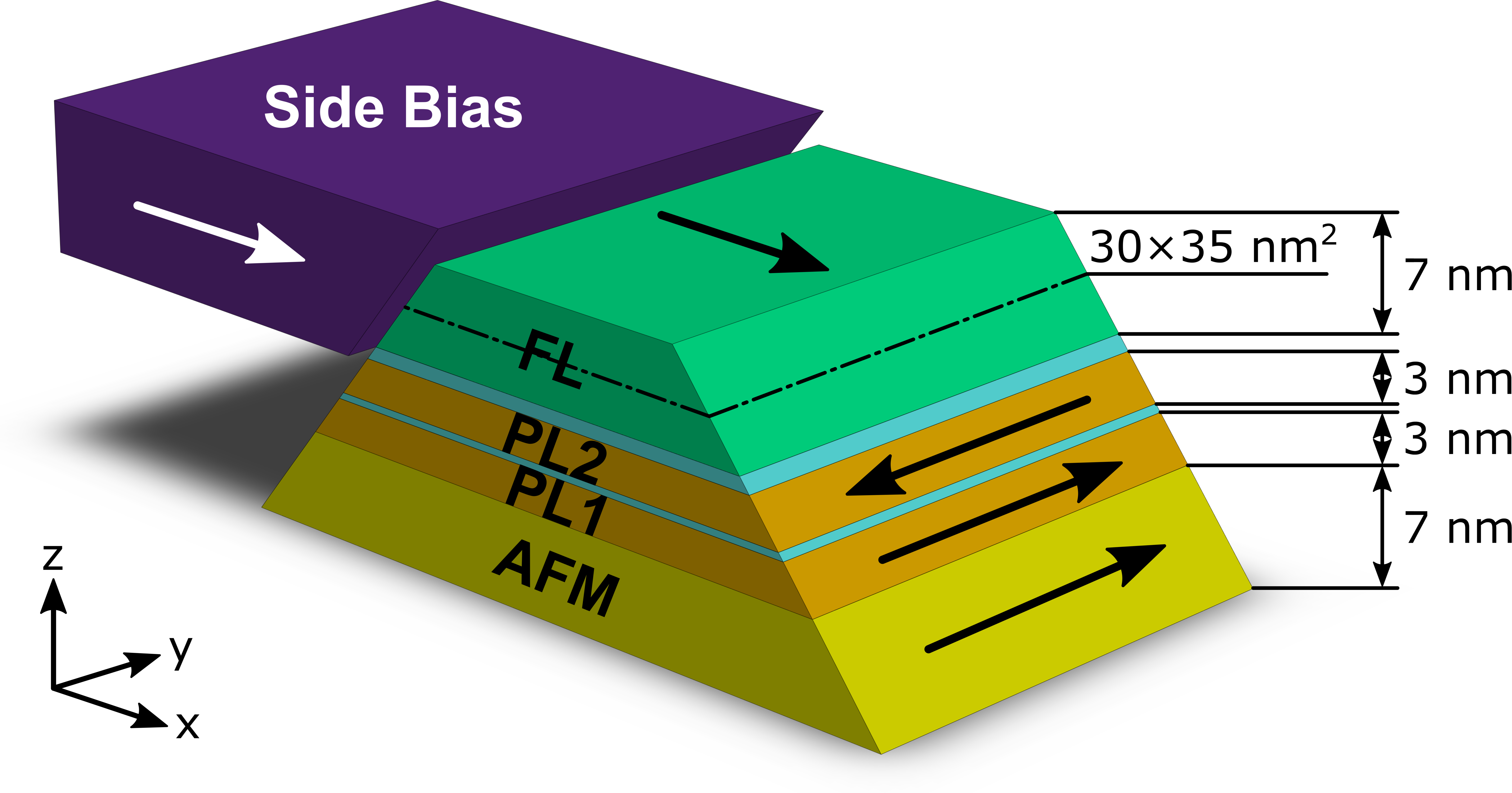}
  \caption{Structure of a state-of-the-art magnetic tunnel junction (MTJ) used in hard disk drives (see text for details). Arrows represent the average magnetization directions of magnetic layers. The right ``Side Bias'' magnet is not shown.}
  \label{fig_magnetic_sensor}
\end{figure}

A magnetic tunnel junction (MTJ) is the basis of modern read sensors commonly used in hard disk drives. As shown in Fig.~\ref{fig_magnetic_sensor}, a typical MTJ represents a nanoscale multilayered structure composed of the free and pinned layers (FL and PL, respectively). The PL is composed of two antiferromagnetically coupled layers (PL1 and PL2) to reduce the stray field from the PL onto the FL. The antiferromagnetic (AFM) layer fixes the magnetization orientation of the PL1 \emph{via} the direct exchange coupling between the AFM and PL1. The side bias field keeps the FL magnetization orthogonal to the PL. The stack's shape is tapered to increase the FL's magnetic stability. The magnetization direction of the PL is assumed to be perfectly fixed while that of the FL rotates in response to the field from the magnetic media. The change in the relative orientation between the FL and PL magnetizations translates into variations in the sensor's tunnel magnetoresistance (TMR), which indicate the difference between the ``up'' and ``down'' bits representing the recorded information~\cite{zhu_2006,taratorin_2011_ch_1}.

MTJs are nonlinear dynamic systems. At room temperature, thermal fluctuations result in random magnetization dynamics of both the FL and PL, which can be quantified by their ferromagnetic resonance (FMR) modes~\cite{kittel_1951}. Another source of magnetization dynamics in MTJs is the spin-torque (ST) effect, which describes a direct transfer of angular momentum from the spin-polarized electrons to the local magnetization~\cite{brataas_2012}. Even in the macrospin approximation, magnetization dynamics are inherently nonlinear because oscillations of magnetization components arise from the precession of the magnetization vector with a constant magnitude, leading to the following nonlinear relation between these components: $m_x^2 + m_y^2 + m_z^2 = 1$~\cite{berkov_2008}. ST-induced magnetization dynamics result in large-angle magnetization precession where the nonlinear nature of magnetization dynamics becomes especially pronounced.

The fundamental theory of nonlinear oscillations states that their nonlinearity results in new resonances such that oscillations of frequency close to $f_0$ (natural frequency) can be excited by an external force with a frequency considerably different from $f_0$. Namely, a resonant condition might occur at every excitation frequency $p f_0 / q$, where $p$ and $q$ are positive integers. In practice, however, $p$ and $q$ should be small because the resonance strength rapidly decreases with increasing order of nonlinearity~\cite{landau_1976_ch_29}.

Specific cases of forced oscillations in nonlinear systems excited at $n f_0$ and $f_0 / n$, where $n$ is a positive integer, are called harmonic (or super-harmonic) and sub-harmonic injection locking, respectively. Injection locking is observed in numerous types of physical systems. Most often, however, this term is associated with electronic oscillators and laser resonators.  In optics, injection locking has been used to improve the frequency stability of lasers and reduce the frequency noise of laser diodes. In electronic systems, injection locking has been used to increase the pull-in (or ``capture'') range and reduce the output phase jitter in phased-locked loops~\cite{maddaloni_2013_ch_2_6_2}.


In ST-driven systems, \emph{e.g.}, spin-torque nano-oscillators (STNOs), super- and sub-harmonic injection locking can be used to generate microwave and millimeter wave signals. In Ref.~\cite{keatley_2016}, Keatley \emph{et al.} excited DC-biased spin-torque vortex oscillators (STVOs) with an AC signal at multiples (harmonics) of their fundamental frequency. The response was measured at a fractional frequency of the excitation signal corresponding to the fundamental frequency. In Ref.~\cite{lebrun_2015}, Lebrun \emph{et al.} presented an experimental study of both super- and sub-harmonic injection locking to an AC excitation signal in DC-biased STVOs. It resulted in pure phase locking with no phase slips and an output power of $>\!\!1~\mu\textrm{W}$ observed at room temperature and zero magnetic field. In Ref.~\cite{carpentieri_2013}, Carpenteri \emph{et al.} performed a numerical study of both super- and sub-harmonic injection locking in STNOs based on hybrid spin-valves composed of two FLs and orthogonal PLs. In Ref.~\cite{quinsat_2011}, Quinsat \emph{et al.} reported the results of a comprehensive synthesis of experimental and numerical studies of the super-harmonic injection locking in STNOs. This work was inspired by the analytical theory of fractional synchronization presented in Ref.~\cite{urazhdin_2010}, which, in turn, concluded that for large AC power levels the coupling between the oscillating magnetic layer and the source becomes nonlinear, resulting in a fractional synchronization regime~\cite{dussaux_2011}.

In this work, we study the sub-harmonic injection locking in MgO-based MTJs. Combining the nonlinear vector network analyzer (NVNA) measurements and micromagnetic simulations, we show that this locking is determined not only by the intrinsic nonlinearity of the magnetization dynamics, but also by the magnetodipolar feedback between the FL and PL. Furthermore, we demonstrate that the orthogonality of magnetization directions of the FL and PL1 also plays an important role, facilitating the read sensors' response at 1/2 the FL FMR frequency $f_\textrm{FL}$. We observe distinct peaks in the DC response at 1/2, 1/3, 1/4, and 1/6 of $f_\textrm{FL}$. Performing corresponding simulations, we show that these resonances are due to the combination of the low-order nonlinearity, magnetodipolar feedback, and the mutual ST effect between the FL and PL1. Importantly, strong magnetodipolar feedback permits sub-harmonic injection locking within a wide range of integer fractions of $f_\textrm{FL}$.

The presented work suggests that NVNA measurements in conjunction with micromagnetic modeling can greatly assist in clarifying the uncertainty in defining the system's magnetic and ST parameters. Furthermore, such a synthesis of nonlinear measurements and modeling identified the need to consider mutual ST between the FL and PL in micromagnetic modeling.

\section{\label{sec_nonlinear_characterization}Nonlinear characterization}

We use the advances in microwave measurement techniques to determine the read sensors' DC and harmonic responses, which comprise nonlinear characterization. Due to the broadband nature of and variability in the GHz-range FMR in read sensors, the nonlinear characterization was performed over a wide frequency range (the excitation frequency ranging from 1 to 15 GHz). Additionally, due to the presence of magnetic shields and bias magnets (Fig.~\ref{fig_magnetic_sensor}), our samples are unsuitable for characterization with an applied magnetic field. The shields would distort the external field such that one would be unable to control the relative orientation of the FL and PLs to the degree required for such studies. Therefore, all measurements presented in this work were performed without an applied magnetic field, and the excitation was a combination of AC power and DC bias current.

\subsection{\label{subsec_DC_response}DC response}

We start with the definition of the voltage induced in the read sensor, which is subjected to both DC and AC currents:

\begin{equation}
    I = I_\textrm{DC} + I_\textrm{AC}\sin{(\omega t)}.
\end{equation}

In the presence of the TMR effect, the system's resistance has two constituents: mag\-ne\-ti\-za\-tion-independent ohmic resistance $R_0$ and magnetization-dependent, TMR-caused resistance $R_\textrm{TMR}(t)$. Hence, the resultant time-dependent voltage response is

\begin{equation}
    V(t) = I(t) R(t) = \left[ I_\textrm{DC} + I_\textrm{AC}\sin{(\omega t)} \right] \times \left[ R_0 + R_\textrm{TMR}(t) \right].
    \label{eq_oscillating_voltage_response}
\end{equation}

If $I_\textrm{DC}$ is less than the critical current above which the ST-induced steady-state oscillations arise, the only excitation present in the system is the periodic AC current. Thus, the time-dependent magnetization resulting in the time-dependent TMR term is a periodic function with a period $T = 2 \pi / \omega$. Due to various sources of nonlinearities in the system, the TMR-caused resistance can be non-harmonic and subsequently be expressed as a Fourier series:

\begin{equation}
    R_\textrm{TMR}(t) = R_\textrm{TMR}^\textrm{av} + \sum_{n = 1}^\infty a_n \cos{(n \omega t)} + \sum_{n = 1}^\infty b_n \sin{(n \omega t)},
    \label{eq_TMR_term_Fourier_series}
\end{equation}

\noindent where the Fourier coefficients $a_n$ and $b_n$ have the dimensionality of resistance. The averaged $R_\textrm{TMR}^\textrm{av}$ is determined by the nonlinearity of the magnetization oscillations and is highly sensitive to ST asymmetry. This issue is discussed in more detail in Section~\ref{sec_results_and_discussion}. Substituting Eq.~(\ref{eq_TMR_term_Fourier_series}) into Eq.~(\ref{eq_oscillating_voltage_response}) and then time-averaging the result produces the following DC voltage response:

\begin{equation}
    V_\textrm{DC} = \langle V(t) \rangle = \underbrace{\vphantom{\frac{I_\textrm{AC} b_1}{2}}I_\textrm{DC} R_0 + I_\textrm{DC} R_\textrm{TMR}^\textrm{av}}_{\textrm{``constant''}} + 
    \underbrace{I_\textrm{AC} \frac{b_1}{2}}_{\textrm{``oscillating''}},
    \label{eq_dc_voltage_response}
\end{equation}

\noindent where all other higher-order terms disappear after time-averaging due to the orthogonality properties of harmonic functions.

In Eq.~(\ref{eq_dc_voltage_response}), the ``constant'' term comes from the magnetization-independent (ohmic resistance) and time-averaged magnetization-dependent, TMR-caused resistances. The ``oscillating'' term involves mixing the excitation signal with the TMR oscillations only at the same frequency~\cite{sankey_2006}. From Eq.~(\ref{eq_dc_voltage_response}), it can also be seen that the contribution from the ``oscillating'' term depends on the phase difference $\Delta \phi$ between these two signals: its magnitude is at its maximum at $\Delta \phi = 0^{\circ}$ or $180^{\circ}$ (in- or out-of-phase) and its minimum at $\Delta \phi = 90^{\circ}$ or $270^{\circ}$ (quadrature).

\begin{figure}[!htb]
  \centering
    \includegraphics[scale=.75]{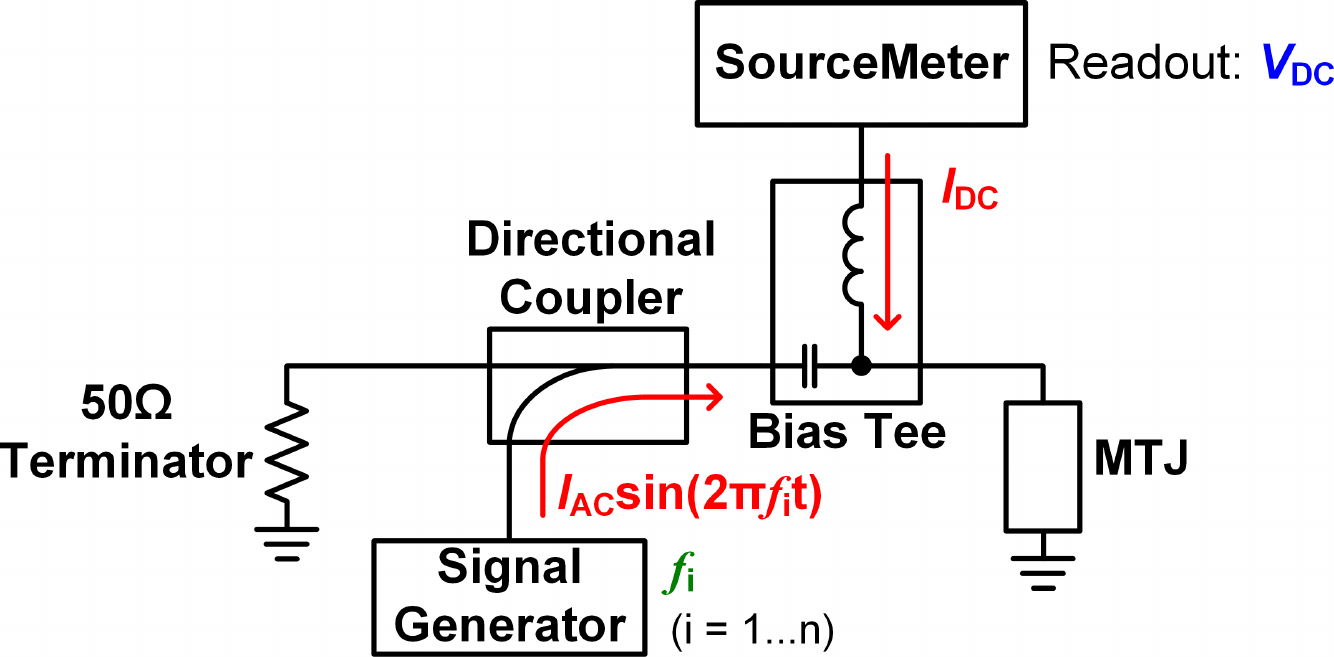}
  \caption{DC readout ferromagnetic resonance (FMR) measurement relying on the spin-torque (ST) rectification phenomenon.}
  \label{fig_dc_readout}
\end{figure}

In this study, the typical MTJ's dimensions are $30\times35~\textrm{nm}^2$ in the $x$-$y$ plane and $\approx~\!\!25~\textrm{nm}$ in the $z$-direction. Its resistance-area product and TMR ratio are $0.4~\Omega \mu \textrm{m}^2$ and $90\%$, respectively. Experimentally, the sensor's DC response can be extracted using the steady-state DC readout measurement technique~\cite{tulapurkar_2005,sankey_2006}. In the simplest measurement configuration (Fig.~\ref{fig_dc_readout}), the MTJ-based read sensor is excited with an AC signal for different DC bias currents, while the DC voltage across the sensor is measured with a sourcemeter.

\begin{figure}[!htb]
  \centering
    \includegraphics[scale=.35]{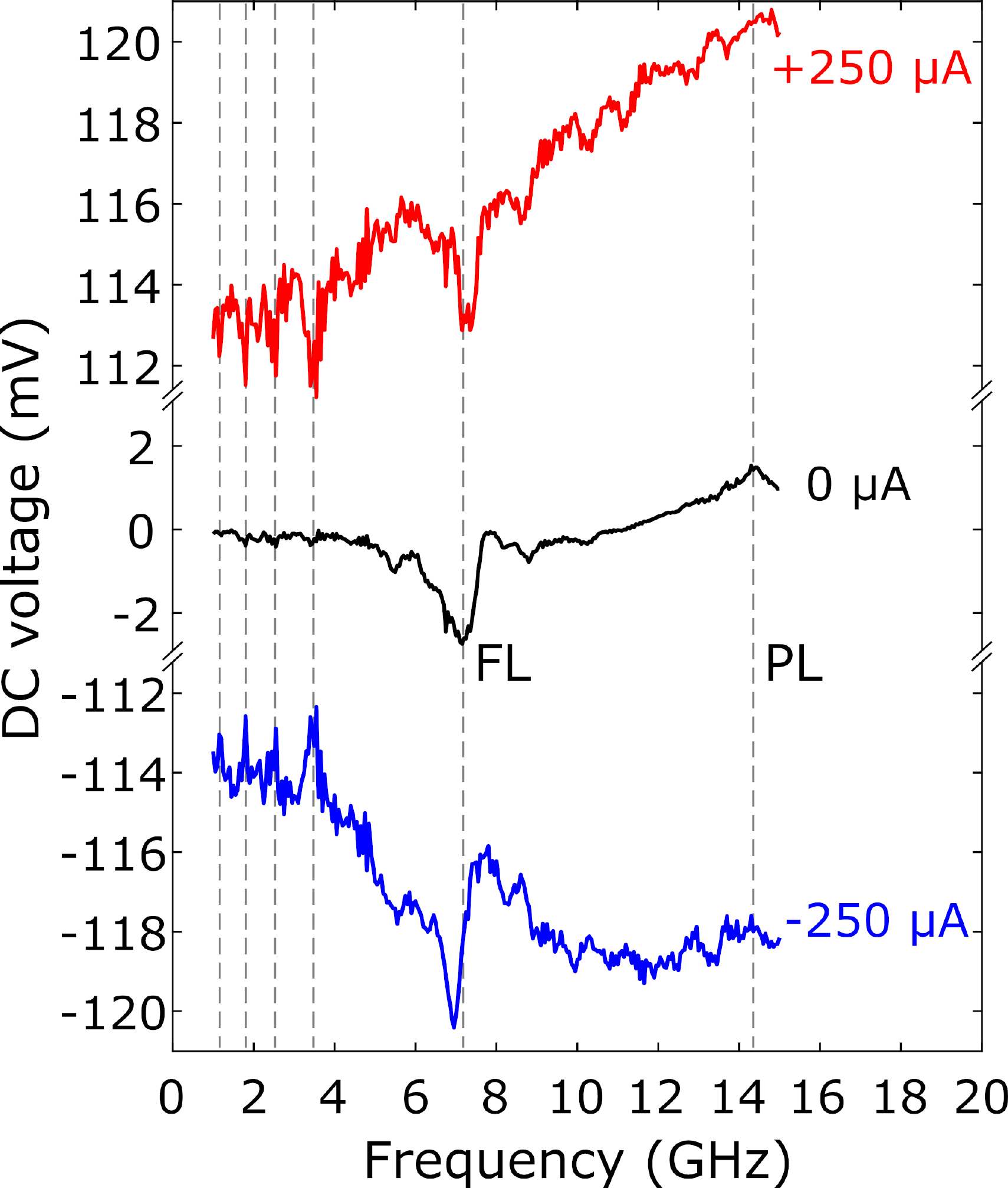}
  \caption{DC response as a function of the excitation frequency obtained at $-5~\textrm{dBm}$ source power and different DC bias currents. The relatively high excitation level of $-5~\textrm{dBm}$ was selected to emphasize the nonlinear phenomenon. ``FL'' and ``PL'' denote the free and pinned layers' natural resonant precession frequencies~\footnote{These frequencies are consistent with those obtained with the thermal noise FMR (T-FMR) measurement under the same DC biasing condition.}. Positive DC bias corresponds to the electron flow from the PL to FL.}
  \label{fig_dc_readout_-5_dBm}
\end{figure}

In this study, six Western Digital production read sensors were characterized.  The representative sample's DC readout measurements shown in Fig.~\ref{fig_dc_readout_-5_dBm} were obtained using the measurement technique introduced in Fig.~\ref{fig_dc_readout} (utilizing a Keysight PNA-X N5247A as the AC source). Our main observation is the presence of peaks emerging in the DC readout measurements with non-zero DC bias current at frequencies that are the integer fractions ($1/2$, $1/3$, $1/4$, and $1/6$) of the FL's resonant precession frequency $f_\textrm{FL}$. A similar observation was made in a micromagnetic study of the current-perpendicular-to-plane (CPP) spin valve heads under ST excitation~\cite{zhu_2004}. A subsequent study of ST-induced magnetization dynamics in thin magnetic nanoelements demonstrated that nanoelements of a certain size can exhibit splitting of their magnetization precession trajectory into limiting sub-cycles~\cite{berkov_gorn_2005}. In the spectrum, these sub-cycles corresponded to peaks at frequencies lower than that of the complete motion cycle. In Ref.~\cite{berkov_gorn_2005}, this regime precedes the state of chaos characterized by strongly inhomogeneous large-angle magnetization precession, which produces chaotic trajectories and almost continuous spectrum with no distinct FMR peaks. In both micromagnetic studies, however, the state near bifurcation (characterized by the presence of peaks at the fractional frequencies of $f_\textrm{FL}$) was observed in a \emph{narrow} range of corresponding parameters: applied DC bias currents (2.8 to 3.6~mA) in Ref.~\cite{zhu_2004} or sizes of nanoelements (40 to 52~nm) in Ref.~\cite{berkov_gorn_2005}. In contrast, we observe peaks at the FL's fractional frequencies in a \emph{wide} range of applied AC and DC signals, which requires an alternative explanation.

\subsection{\label{subsec_harm_response}Harmonic response}

The steady-state DC readout measurement technique allows analysis only of the DC spectral component. Since nonlinear effects are involved, a thorough nonlinear harmonic analysis is highly desired. Such an analysis can be accomplished by using an NVNA. Since the characterized magnetic read sensors are one-port devices, this analysis is limited to the one-port scenario.

\begin{figure}[!htb]
  \centering
  \savebox{\largestimage}{\includegraphics[scale=.4]{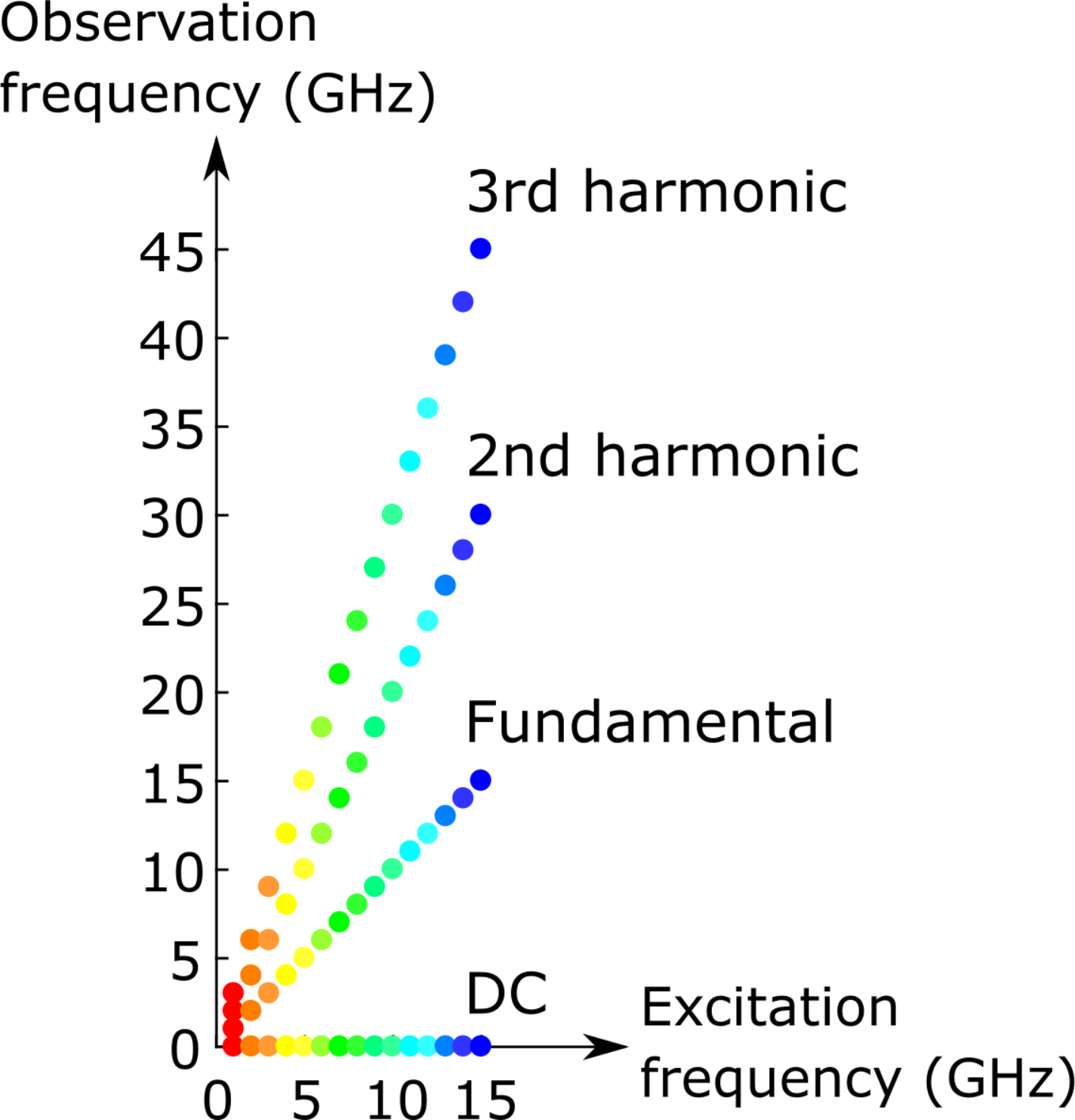}}%
  \begin{subfigure}[b]{.31\linewidth}
    \centering
    \raisebox{\dimexpr.5\ht\largestimage-.5\height}{%
      \includegraphics[scale=.6]{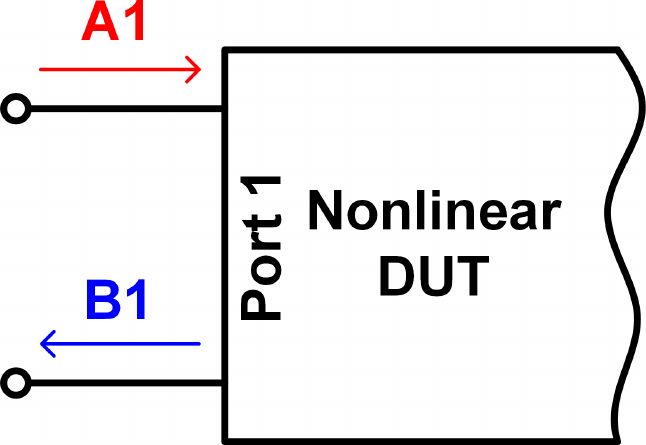}}
    \caption{}
    \label{subfig_nonlinear_dut}
  \end{subfigure}
  \enskip
  \begin{subfigure}[b]{.31\linewidth}
    \centering
    \raisebox{\dimexpr.5\ht\largestimage-.5\height}{%
      \includegraphics[scale=.7]{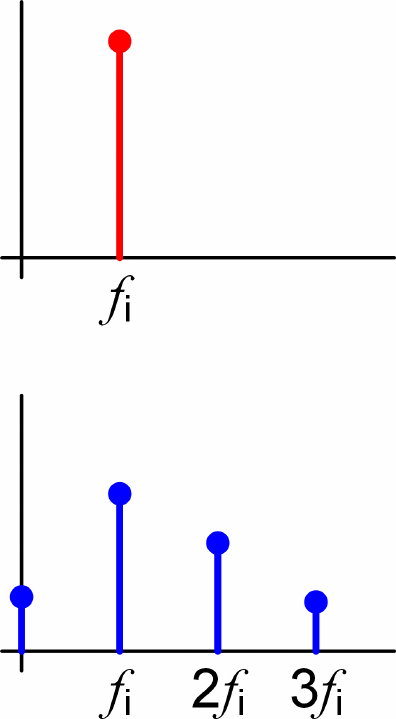}}
    \caption{}
    \label{subfig_nonlinear_response}
  \end{subfigure}
  \enskip
  \begin{subfigure}[b]{.31\linewidth}
    \centering
    \usebox{\largestimage}
    \caption{}
    \label{fig_nvna_freq_sweep}
  \end{subfigure}
  \caption{Nonlinear vector network analyzer (NVNA) measures the incident and reflected waves' ($A_1$ and $B_1$, respectively) magnitudes and phases (\subref{subfig_nonlinear_dut}) at the excitation frequency and at harmonic components (\subref{subfig_nonlinear_response}) to which energy may be transferred due to the device under test's (DUT's) nonlinear characteristics. (\subref{fig_nvna_freq_sweep}) shows the corresponding excitation and observation frequencies.}
  \label{fig_nvna_measurement}
\end{figure}

\begin{figure}[!htb]
\centering

\begin{subfigure}{.49\linewidth}
\centering
\includegraphics[scale=.35]{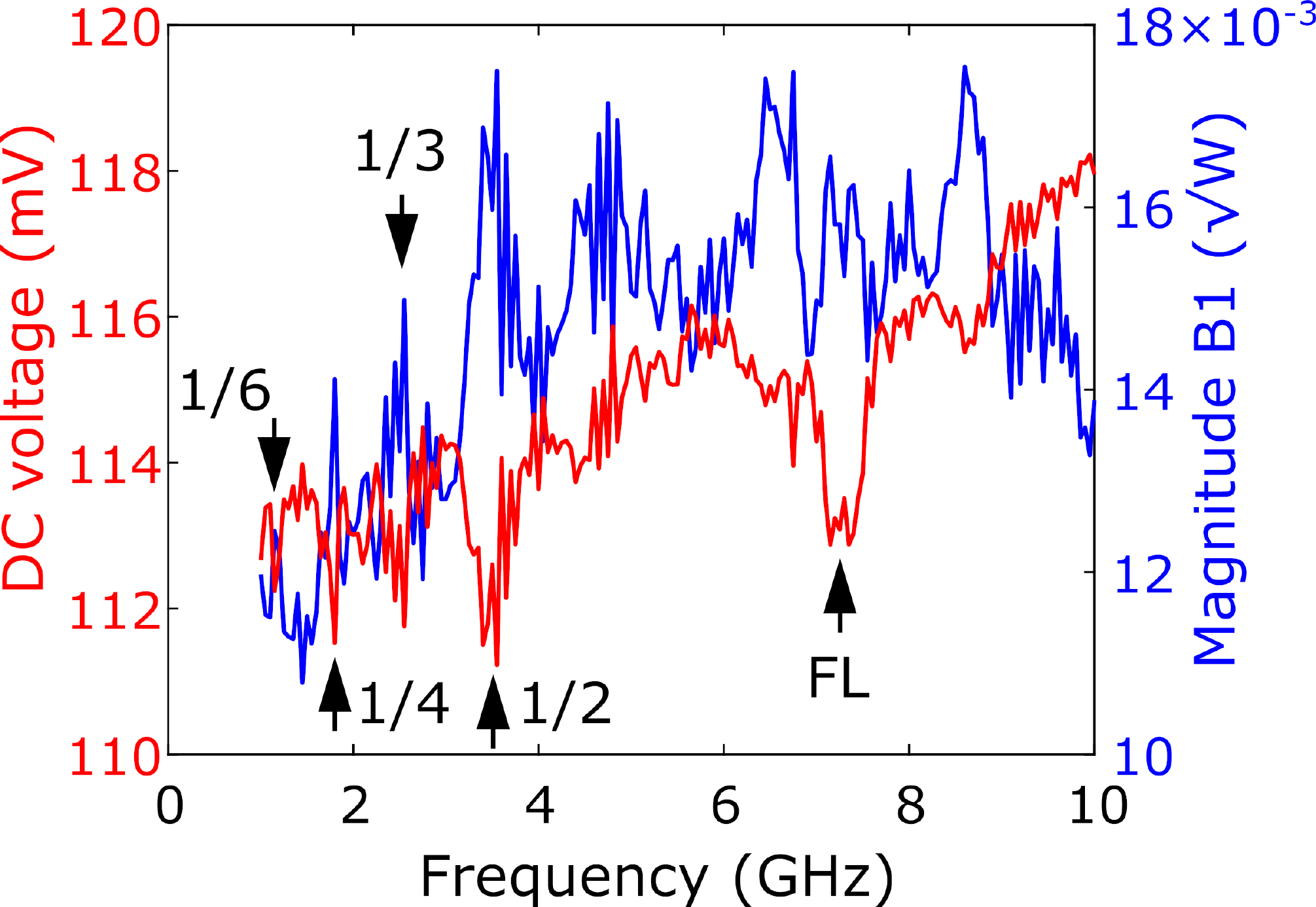}\hfill
\caption{fundamental}\label{fig_1st_harm_-5_dBm_+250_uA}
\end{subfigure}
\begin{subfigure}{.49\linewidth}
\centering
\includegraphics[scale=.35]{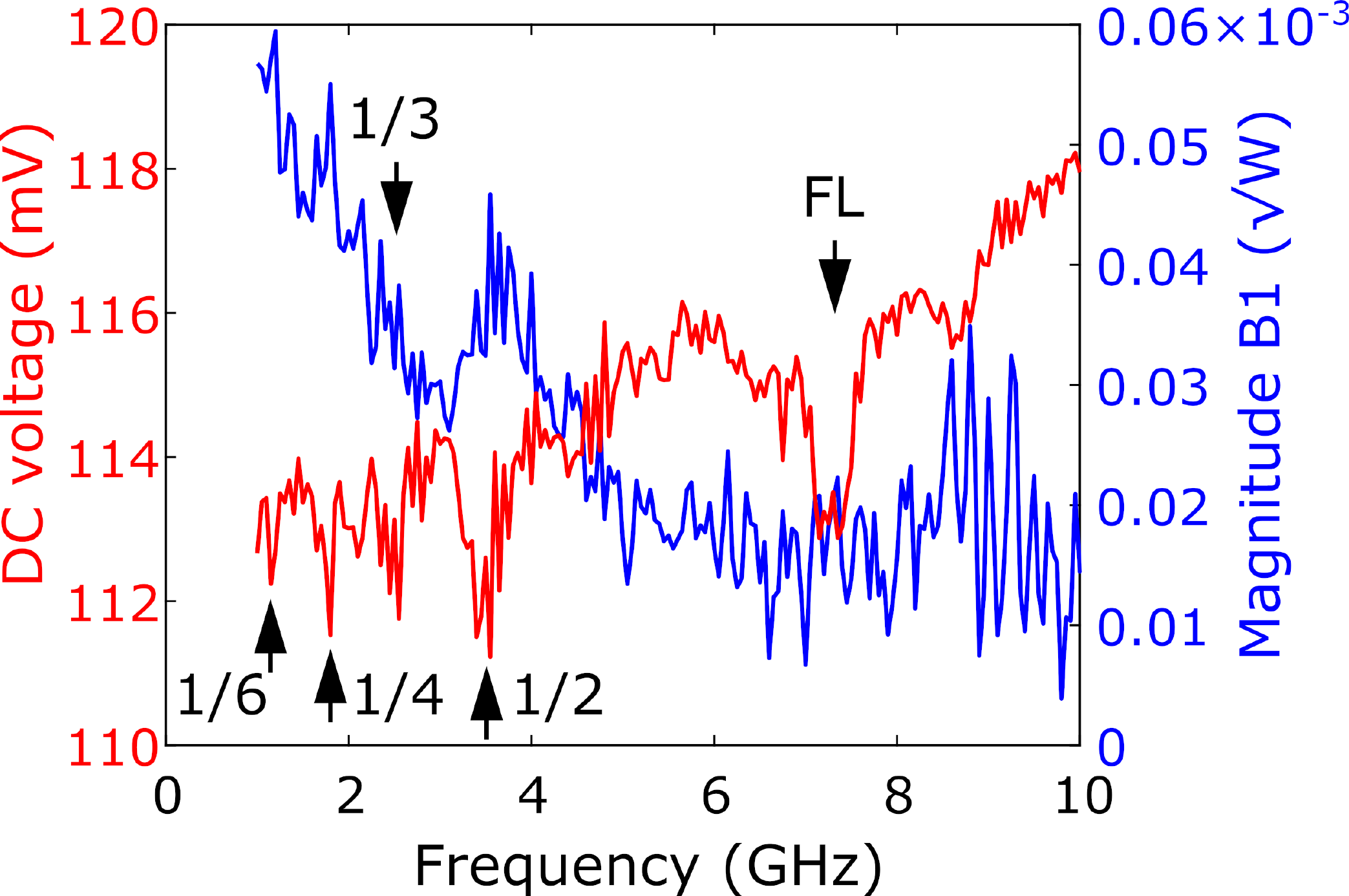}
\caption{2nd harmonic}\label{fig_2nd_harm_-5_dBm_+250_uA}
\end{subfigure}

\medskip

\begin{minipage}{.48\linewidth}
\caption{Representative sample's harmonic response (blue curves) measured with a Keysight PNA-X N5247A at $+250~\mu\textrm{A}$ DC bias current, $-5~\textrm{dBm}$ AC power, and 15~Hz intermediate frequency bandwidth (IFBW). (\subref{fig_1st_harm_-5_dBm_+250_uA}) fundamental, (\subref{fig_2nd_harm_-5_dBm_+250_uA}) second, and (\subref{fig_3rd_harm_-5_dBm_+250_uA}) third harmonics of $B_1$ plotted \emph{versus} the excitation frequency. The DC response is shown in red.}
\label{fig_harmonic_response_-5_dBm_+250_uA}
\end{minipage}\hfill
\begin{subfigure}{.49\linewidth}
\centering
\vspace{0pt}
\includegraphics[scale=.35]{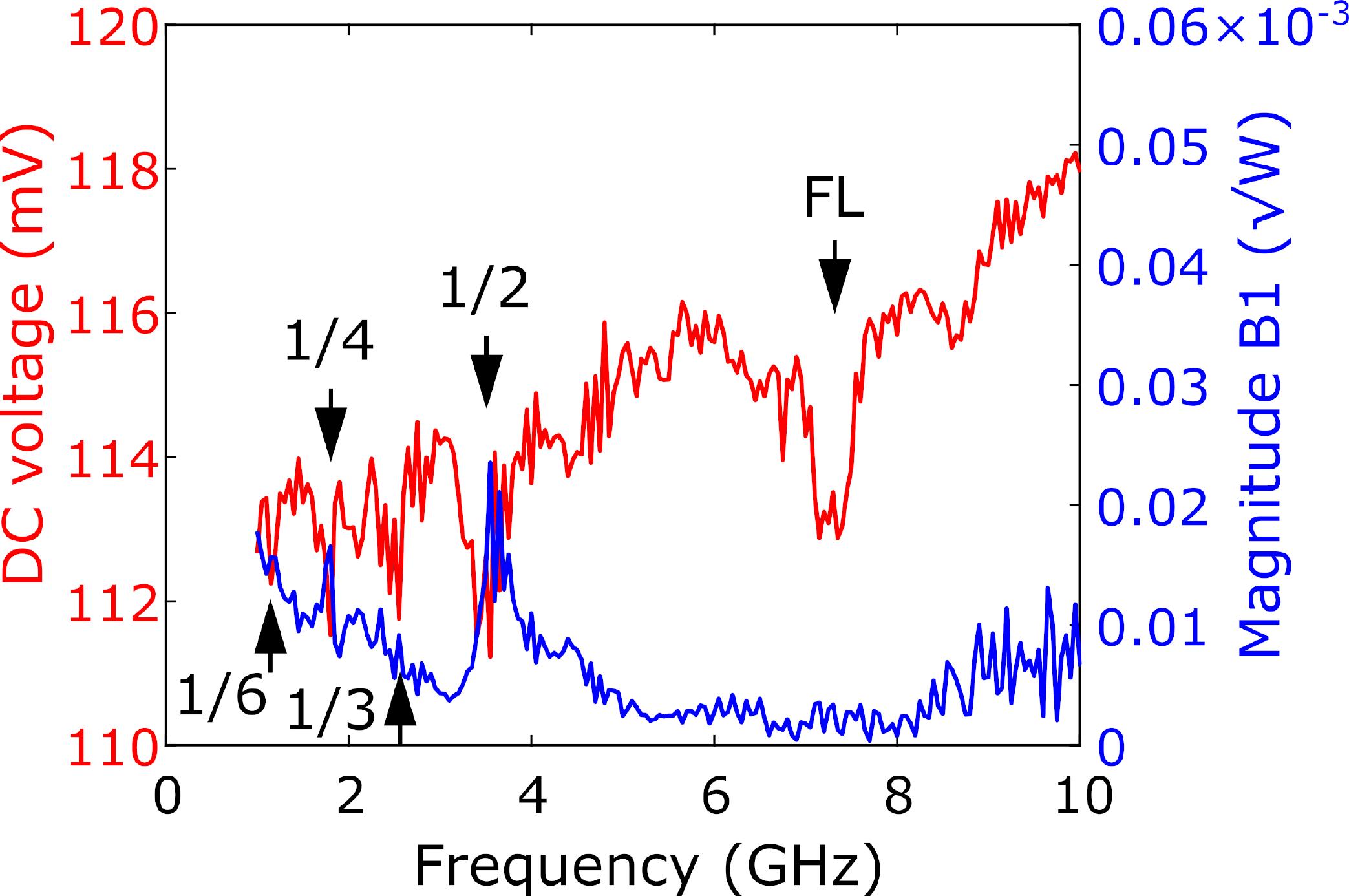}
\caption{3rd harmonic}\label{fig_3rd_harm_-5_dBm_+250_uA}
\end{subfigure}

\end{figure}

In contrast to the \emph{linear} VNA, which measures the magnitude ratio and phase difference between the incident and reflected waves ($A_1$ and $B_1$, respectively) only at the excitation frequency, the NVNA measures the actual $A_1$ and $B_1$ waves' magnitudes and phases at the excitation frequency as well as harmonic components to which energy may be transferred due to the device's nonlinear characteristics (Fig.~\ref{fig_nvna_measurement})~\cite{roblin_2011,remley_2009}. In this measurement, the excitation frequency $f_i$ of the incident wave $A_1$ was swept from 1 to 15~GHz. At each frequency, the fundamental of the reflected wave $B_1$ was measured at $f_i$. Naturally, the second harmonic of the reflected wave $B_1$ was measured at $2f_i$ (2 to 30~GHz) and the third harmonic was measured at $3f_i$ (3 to 45~GHz). We emphasize that the resultant measurements represent the frequency response of $B_1$, not the power spectrum. A power spectrum is the distribution with frequency of the power content of the signal~\cite{randall_1987_ch_2_2_1}, whereas the frequency response~\cite{ziemer_2014_ch_2_6_4} curves in Fig.~\ref{fig_harmonic_response_-5_dBm_+250_uA} represent the magnitude of the corresponding harmonic component of $B_1$ as a function of the excitation frequency and are linearly proportional to voltage.

In Fig.~\ref{fig_harmonic_response_-5_dBm_+250_uA}, the magnitudes of the fundamental, second, and third harmonics of $B_1$ are counterposed to the DC readout curve. In Fig.~\ref{fig_1st_harm_-5_dBm_+250_uA}, the fundamental of $B_1$ exhibits distinct peaks at frequencies corresponding to $f_\textrm{FL}$, $f_\textrm{FL}/2$, $f_\textrm{FL}/4$, and (less pronounced) $f_\textrm{FL}/6$. These peaks have their counterparts on the DC response curve. In addition, strong peaks are observed in-between $f_\textrm{FL}$ and $f_\textrm{FL}/2$. They might correspond to resonances at frequencies $p f_\textrm{FL}/q$ (where $p$ and $q$ are positive integers) and have no analogues in the DC response. Understanding the nature of these additional peaks requires further investigation. In Figs.~\ref{fig_2nd_harm_-5_dBm_+250_uA} and~\ref{fig_3rd_harm_-5_dBm_+250_uA}, the second and third harmonics of $B_1$ also reveal distinct peaks at $f_\textrm{FL}/2$, $f_\textrm{FL}/4$, and (less pronounced) $f_\textrm{FL}/6$. Interestingly, the FL mode and the peak located at $f_\textrm{FL}/3$ \emph{do not} produce the second and third harmonics of $B_1$.

The NVNA measurements imply that the read sensor's nonlinear characteristics give rise to nonlinear oscillations under the AC excitation signal with the frequency considerably lower than, but still an integer ratio of, the FL mode. Naturally, these dynamics also produce a measurable DC contribution corresponding to this excitation frequency.


\section{\label{sec_micromagnetic_modeling}Micromagnetic modeling}

We employ micromagnetic modeling to understand the underlying physics that enables the DC and harmonic responses at frequencies corresponding to the FL FMR mode's fractional ratios. Due to the model's complexity and uncertainty in the definition of some of its magnetic and ST parameters, we find it vital to discuss in detail the modeling methodology and modeling considerations.

\subsection{\label{sec_modeling_methodology}Modeling methodology}

All numerical simulations presented in this work were performed using the micromagnetic software MicroMagus, which solves the modified Landau-Lifshitz-Gilbert (LLG) equation for the magnetization $\textrm{\bf{M}}$~\cite{micromagus}

\begin{equation}
    \dv{\bf{M}}{t} = -\gamma 
    \left[
    \bf{M} \cross \left( \bf{H}_{\rm det} + \bf{H}_{\rm th} \right) 
    \right] 
    - \gamma \frac{\lambda}{M_s} 
    \left[ 
    \bf{M} \cross \left[ \bf{M} \cross \left( \bf{H}_{\rm det}  + \bf{H}_{\rm th} \right) \right] \right],
    \label{eq_modified_llg}
\end{equation}

\noindent using one of the optimized Runge-Kutta or Bulirsch-Stoer algorithms with an adaptive step-size control for both $T = 0$ and $T >0$. In Eq.~(\ref{eq_modified_llg}), the precession constant $\gamma$ is defined \emph{via} the absolute value of the gyromagnetic ratio $\gamma_0$ as $\gamma = \gamma_0/(1 + \lambda^2)$. The damping constant $\lambda$ is equal to the corresponding damping $\alpha$ in the LLG form where the magnetization derivative is present on both sides of the equation.

The deterministic field $\bf{H}_{\rm det}$ contains four standard contributions: external, anisotropy, exchange, and magnetodipolar interaction fields. In our case, the ST effect is taken into account \emph{via} an additional effective field term, which in the standard Slonczewski formalism
~\cite{slonczewski_1996,slonczewski_2002} has the form

\begin{equation}
    {\bf H}_{\rm ST} = f_J(\theta) \left[ {\bf M} \cross {\bf p} \right].
    \label{eq_field-like_term}
\end{equation}

Here, the dimensionless ST amplitude $f_J$ depends on the angle $\theta$ between the magnetization $\textrm{\bf{M}}$ and the spin-polarization direction $\textrm{\bf{p}}$ as follows~\cite{slonczewski_2002,xiao_2004}:

\begin{equation}
    f_J(\theta) = a_J \frac{2 \Lambda^2}{(\Lambda^2 + 1) + (\Lambda^2 - 1) \cos{\theta}},
    \label{eq_st_amplitude}
\end{equation}

\noindent where the factor $a_J$ is given by

\begin{equation}
    a_J = \frac{\hbar}{2 \abs{e}} \frac{j P}{M_s^2 d}.
    \label{eq_reduced_st_amplitude}
\end{equation}

In Eqs.~(\ref{eq_st_amplitude}) and~(\ref{eq_reduced_st_amplitude}), $e$ is the electron charge, $j$ is the electric current density, and $d$ is the thickness of the magnetic layer subjected to ST. The asymmetry parameter $\Lambda$ strongly depends on the sample configuration and various transport coefficients~\cite{slonczewski_2002,xiao_2004}. When $\Lambda = 1$, the ST effect is assumed to be symmetric. $P$ is the degree of spin polarization of the electrical current~\footnote{To describe the ST's angular dependence in MgO-based junctions, we used the common formulations for metallic junctions. Recent work (Ref.~\cite{kowalska_2018}) shows that the effective difference in angular dependence for these two scenarios is insignificant.}.

In the LLG equation (\ref{eq_modified_llg}), thermal effects are accounted for by the thermal field term $\bf{H}_{\rm th}$ describing random fluctuations induced by the interaction of the ferromagnet with the thermal bath. Components of this fluctuation field have the following statistical properties:

\begin{equation}
	\begin{split}
		\langle {\bf H}^{\rm th}_{\xi, i} \rangle & = 0, \\
        \langle {\bf H}^{\rm th}_{\xi, i}(0) {\bf H}^{\rm th}_{\psi, j}(t) \rangle & = 2 D \delta(t) \delta_{i j} \delta_{\xi \psi},
        \label{eq_stat_properties}
	\end{split}
\end{equation}

\noindent meaning that these fluctuations are assumed to be uncorrelated in space and time ($i$, $j$ are the discretization cell indices; $\xi,{}\psi = x,{}y,{}z$). The noise power $D$ is proportional to the system temperature $T$:

\begin{equation}
    D = \frac{\lambda}{1 + \lambda^2} \frac{k T}{\gamma \mu},
\end{equation}

Here, $\mu$ is the magnetic moment magnitude of a discretization cell. Unless stated otherwise, all micromagnetic simulations in subsequent sections were performed at $T = 0$~K.

The TMR response introduced in Section~\ref{subsec_DC_response} is the quantitative description of the TMR's dependence on the angle $\theta$ defined between the magnetizations of adjacent layers \cite{jaffres_2001}:

\begin{equation}
    R = \frac{1}{G} = \frac{R_{\perp}}{1 + \frac{\Delta_\textrm{TMR}}{2}\cos \theta},
    \label{eq_tmr}
\end{equation}

\noindent where $R_{\perp}$ is the sensor resistance at the orthogonal state and $\Delta_\textrm{TMR}$ is the TMR ratio.

Equation~(\ref{eq_tmr}) is different from the generally used formulations describing a linear variation of the MTJ's resistance with $\cos \theta$, which appears to be valid only for small values of TMR~\cite{jaffres_2001}.

\subsection{\label{subsec_modeling_considerations}Modeling considerations}

In this study, the MgO-based MTJ devices' PL1, PL2, and FL are made of a similar CoFe/CoFeB alloy. The layers' saturation magnetization and exchange stiffness constant were assumed to be $M_s = 1 \times 10^3~\textrm{G}$ and $A = 1 \times 10^{-6}~\textrm{erg/cm}$, respectively. The intrinsic Gilbert dampings of the FL and PL2 were set to $\lambda_{\rm FL} = \lambda_{\rm PL2} = 0.01$. The IrMn-exchange-pinned damping of the PL1 was initially assumed to be an order of magnitude larger, $\lambda_{\rm PL2} = 0.1$ \cite{smith_2010}. The interlayer coupling strengths were set to values similar to those used in Ref.~\cite{pauselli_2017}: $J_1 = 0.04~\textrm{erg/cm}^2$ between the FL and PL2 (``orange-peel'' coupling) and $J_2 = -1.6~\textrm{erg/cm}^2$ between the PL1 and PL2 (strong antiferromagnetic coupling \emph{via} the Ru interlayer).

For each experimental frequency point at a given AC power level and DC bias, we had to perform an independent time-domain micromagnetic simulation. For linear systems that respond only at the excitation frequency, fast broadband excitation can be accomplished \emph{via} sinc pulses or multifrequency signals (\emph{e.g.}, Schroeder-phased harmonic signals) that are optimal for uniform excitation of all system modes~\cite{schroeder_1970, van_der_ouderaa_1988}. This is done so that, \emph{e.g.}, the AC-susceptibility in the whole frequency range of interest can be sampled in a single simulation run. For nonlinear systems, such an approach is not feasible because 1) nonlinear systems respond not only at the excitation frequency but also at its multiples (\emph{i.e.}, harmonics) which strongly interact with each other and 2) by exciting the system with a strong ST pulse, reaching a dynamic equilibrium state requires significant time, which may be much longer than the duration of the optimal sinc pulse.

To enable faster frequency sweeps around the ranges of interest (\emph{i.e.}, FL FMR frequency and its fractional ratios), the following simplifications to the original read sensor design were adopted: 1) The AFM layer was excluded from simulations, but its effect on the PL1 was taken into account by the corresponding exchange bias field of 1000~Oe~\cite{pauselli_2017}. 2) The left and right side bias magnets were also excluded from simulations. Instead, the side bias demagnetizing field was calculated in the quasi-static solver and then included as an external field in all subsequent dynamic simulations. In this model, each magnetic layer is discretized in-plane into $N_x \cross N_y = 32 \times 20$~cells. No discretization was performed in the out-of-plane direction. We verified that introducing such a discretization even for the thickest layer (\emph{i.e.}, FL) did not lead to any significant changes in final results. With these simplifications, MicroMagus simulations required $\approx\!\!1$~hour to collect 60~nsec of magnetization dynamics at $T = 0$~K and $\approx\!\!15$~hours to simulate 200~nsec at $T = 300$~K.

To the best of the authors' knowledge, there have not been any comprehensive micromagnetic studies on the nonlinear magnetization dynamics used to predict the MTJ's harmonic response. In general, these dynamics and, by extension, its harmonic response should be sensitive to the various magnetic and ST parameters described above. To test this, we excited the read sensor model \emph{via} a combination of AC and DC signals (as in the experiment in Fig.~\ref{fig_harmonic_response_-5_dBm_+250_uA}) at 1/4 the frequency of the FL FMR mode.

\begin{figure}[!htb]
  \centering
  \savebox{\largestimage}{\includegraphics[scale=.45]{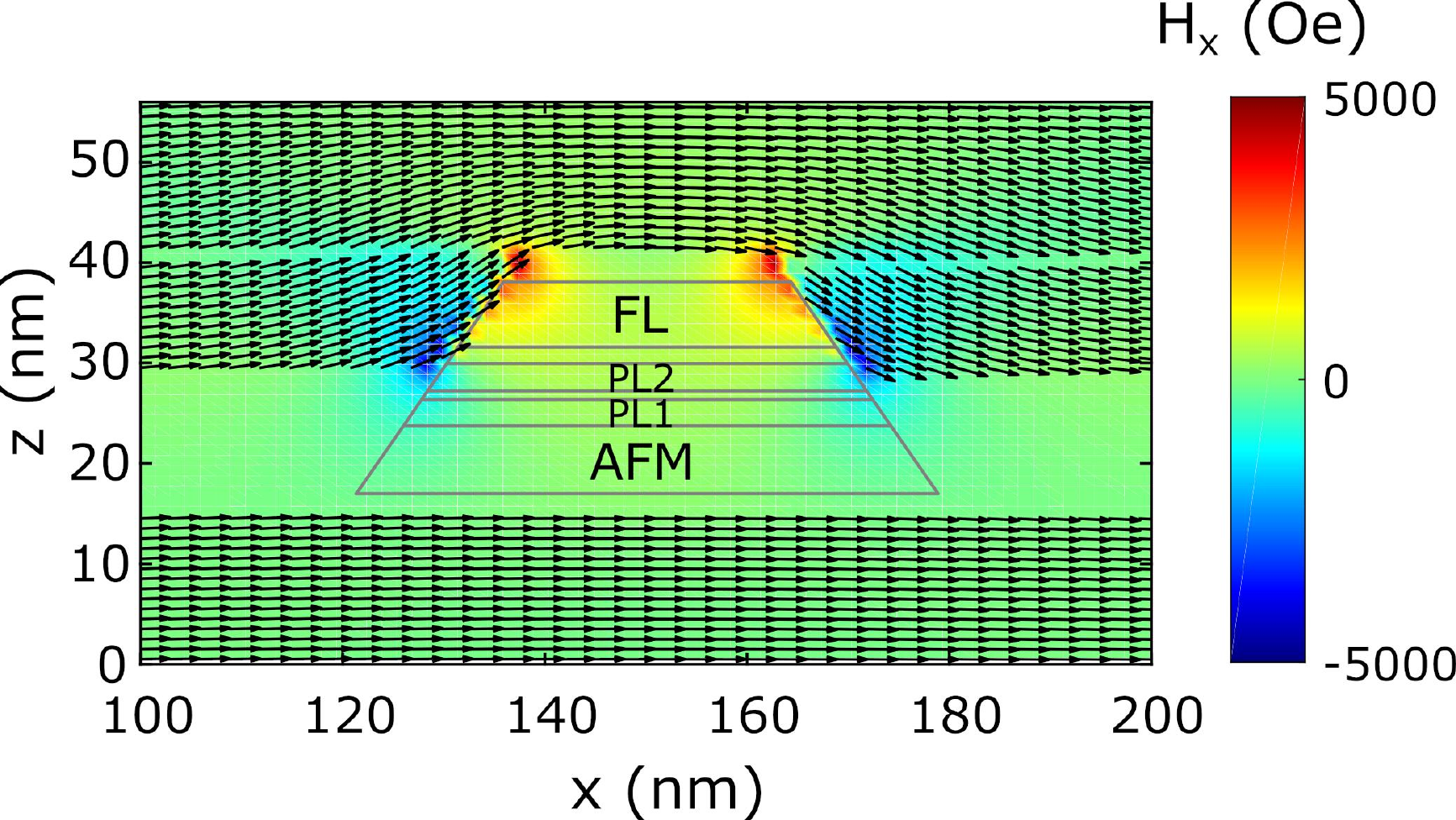}}%
  \begin{subfigure}[b]{.48\linewidth}
    \centering
    \raisebox{\dimexpr.5\ht\largestimage-.5\height}{%
      \includegraphics[scale=.45]{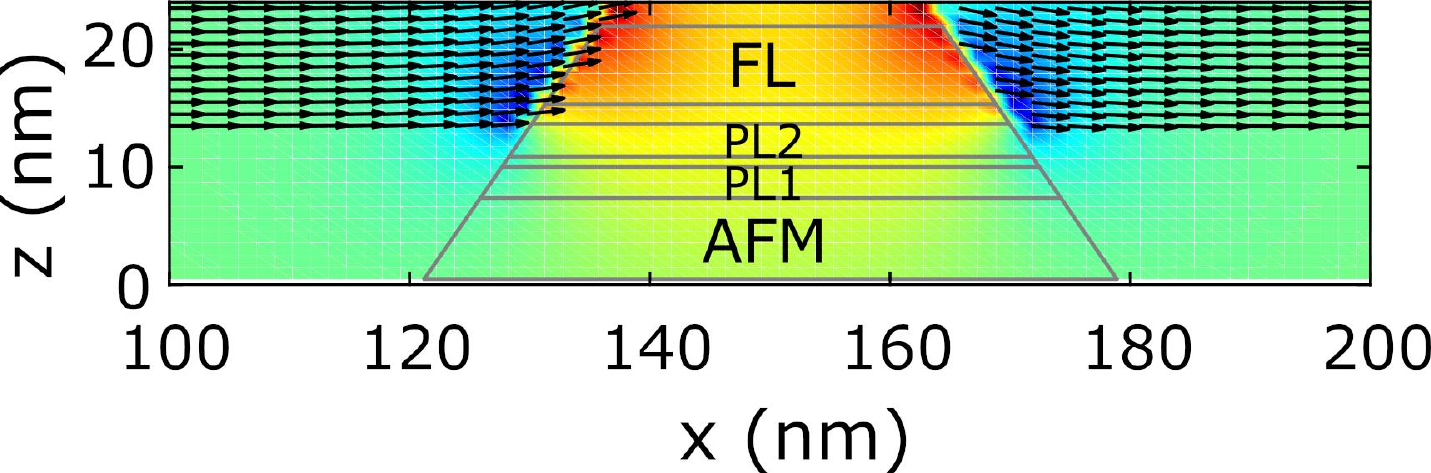}}
    \caption{
    \begin{tabular}{C{1.2cm}|C{1.2cm}|C{1.2cm}}
    \hline
    \multicolumn{3}{c}{Volume-averaged $H_x$ (Oe)}\\
    \hline
    \hline
    FL     &   PL2    &   PL1\\
    \hline
    1829   &   1086   &   559\\
    \hline
    \end{tabular}
    }
    \label{subfig_sb_field_wo_shields}
  \end{subfigure}
  \quad
  \begin{subfigure}[b]{.48\linewidth}
    \centering
    \usebox{\largestimage}
    \caption{
    \begin{tabular}{C{1.2cm}|C{1.2cm}|C{1.2cm}}
    \hline
    \multicolumn{3}{c}{Volume-averaged $H_x$ (Oe)}\\
    \hline
    \hline
    FL     &   PL2    &   PL1\\
    \hline
    861    &   310    &   17\\
    \hline
    \end{tabular}
    }
    \label{subfig_sb_field_w_shields}
  \end{subfigure}
  \caption{Simulated side bias field (\subref{subfig_sb_field_wo_shields}) without and (\subref{subfig_sb_field_w_shields}) with the presence of magnetic shields. The corresponding tables contrast the side bias field onto the free and pinned layers (FL and PL2/PL1, respectively) for these two scenarios. The arrows represent the side bias' and shields' magnetic moments.}
  \label{fig_side_bias_field}
\end{figure}

\emph{Effect of the side bias and magnetic shielding.} In Fig.~\ref{fig_magnetic_sensor}, the read sensor model is shown without the magnetic shields. In a real system, the top and bottom magnetic shields isolate the sensor from adjacent bits and large writer fields. Material parameters of both shields are typical of permalloy. The shields are sufficiently larger than the read sensor. Being unable to micromagnetically model the full-size shields, we reduced their dimensions to $L_x \cross L_y \cross L_z = 300 \times 36 \times 100~\textrm{nm}^3$ and applied the periodic boundary conditions along the $x$-axis to avoid the influence of artificial magnetic ``surface charges'' from the vertical (in the $y$-$z$ plane) shield surfaces. By doing this, we estimated the quasi-quantitative effect of these shields on the side bias field (Fig.~\ref{fig_side_bias_field}).

Figure~\ref{fig_side_bias_field} shows that in the presence of magnetic shields, the simulated side bias field, both onto the FL and especially onto the pinned layers PL2/PL1, sufficiently decreased. This means that the total side bias field (\emph{i.e.}, taking into account the presence of magnetic shields) strongly depends on the shield configuration. Additionally, variations in the spacing between the side bias and the MTJ stack within the fabrication tolerances may affect the side bias field strength. The side bias magnetization is also known to be accurate within approximately $\pm 10\%$. Hence, without having more accurate information concerning the geometry and magnetic parameters of the shields and side bias, we have to adjust the side bias field to match the most reliable experimental results.

The most direct experimental observation is the measured FL FMR frequency. This value can be used to adjust the total external field onto the FL. Based on the obtained value of $f_\textrm{FL} \approx 7~{\rm GHz}$, we have determined that the increase in the simulated side bias field onto the FL by a factor of $\approx\!\!1.7$ is necessary to reproduce this frequency. Moreover, we have found that the proposed adjustment of the side bias field onto the FL has shifted the system's dynamic regime from quasi-chaotic towards pure phase-locking when excited at fractional frequencies. On one hand, the quasi-chaotic regime results in a stronger DC response. On the other hand, this regime was observed only within a narrow range of magnetic and ST parameters and thus is most probably absent in our system. Hence, the quasi-chaotic regime should be avoided, which also justifies the increase in the FL side bias field.

\begin{figure}[!htb]
\centering
\subcaptionbox{$0.46 \times H_x$\label{subfig_H_SB=0_46}}%
  [.49\linewidth]{\includegraphics[scale=.35]{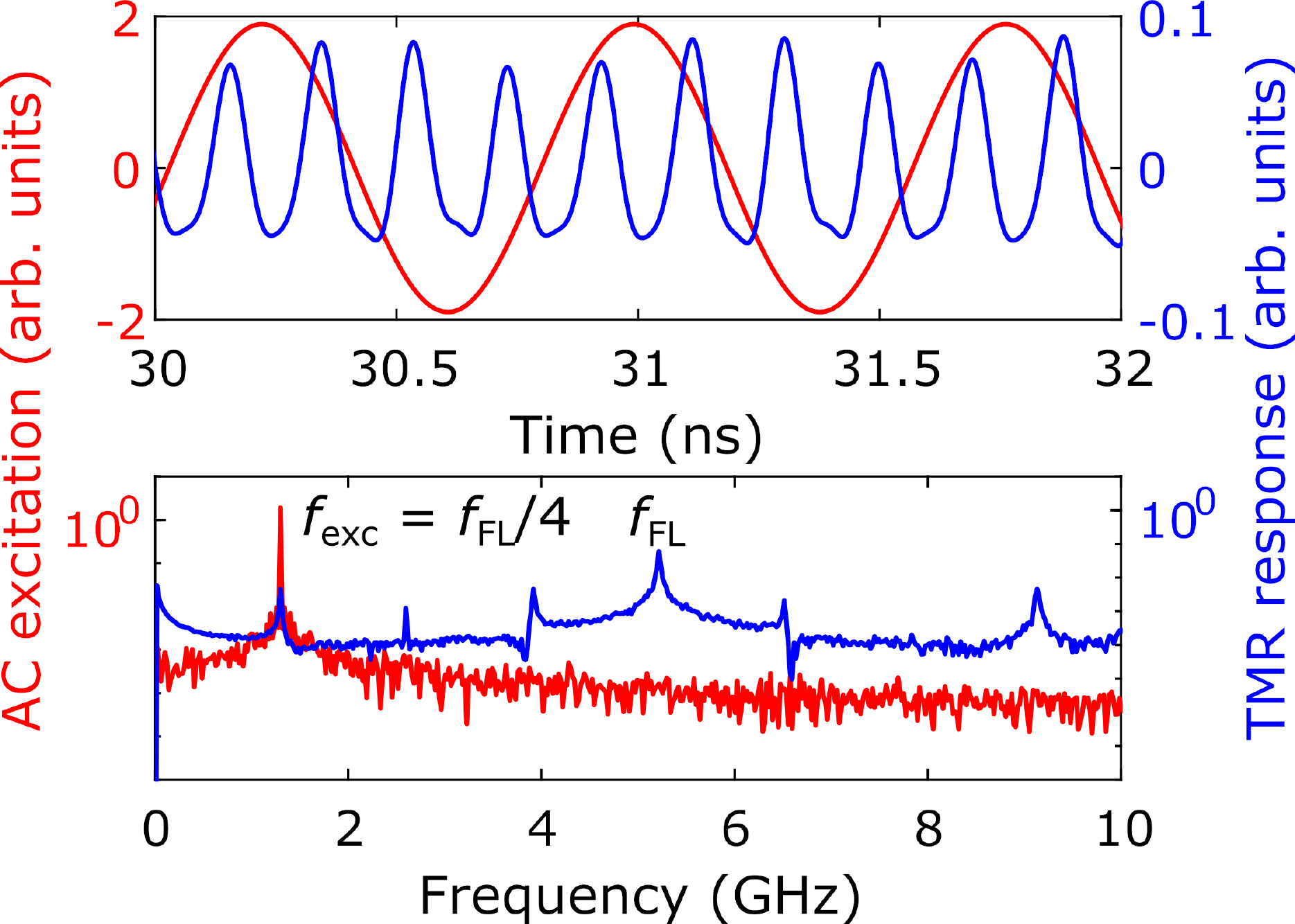}}\hfill
\subcaptionbox{$H_x$\label{subfig_H_SB=1_00}}
  [.49\linewidth]{\includegraphics[scale=.35]{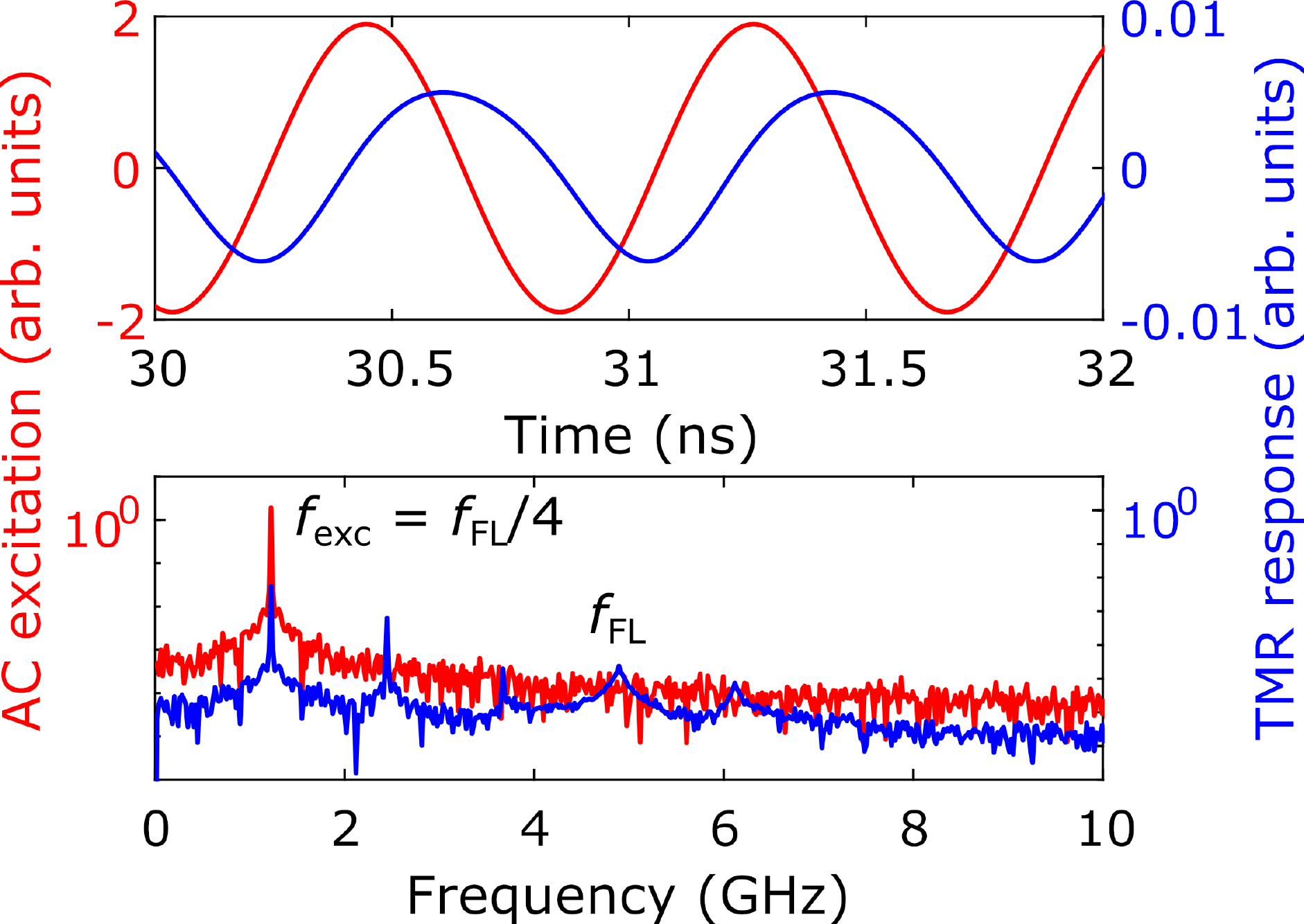}}
\caption{Time- (top) and frequency-domain (bottom) responses of the read sensor model excited at 1/4 the free layer (FL) ferromagnetic resonance (FMR) frequency $f_\textrm{FL}$. Lower side bias field onto PL2/PL1 (\subref{subfig_H_SB=0_46}) evokes strongly nonlinear magnetization oscillations accompanied by stronger harmonic response than in (\subref{subfig_H_SB=1_00}). In the amplitude spectra, note the ordinate axes' logarithmic scale.}
\label{fig_H_SB_f_exc=f_0div4}
\end{figure}

Adjustment of the side bias field onto the pinned layers PL2/PL1 is a more subtle issue. Here, our main criterion was that the side bias field onto the pinned layers should be such that it allows strongly nonlinear magnetization dynamics, leading to the generation of the higher-order harmonics and the sought after DC response (Fig.~\ref{fig_H_SB_f_exc=f_0div4}). Performing test simulations, we discovered that a factor of $\approx\!\!0.2$ reduction in the simulated side bias field onto the PL2/PL1 is necessary to obtain the measurable DC response and higher-order harmonics.

\emph{Effect of damping.} Another ambiguity is the value of the effective damping $\lambda$ included in the LLG equation used for simulations. A lower damping constant corresponds to lower power absorption by the magnetic system, consequently leading to nonlinear oscillations with a larger amplitude and, by extension, a stronger harmonic response.  While 0.01--0.02 is a typical damping constant for the FL and PL2~\cite{smith_2001}, the PL1 damping was initially set to a much higher value, $\lambda_{\rm PL1} = 0.1$~\cite{smith_2010}. In Ref.~\cite{smith_2010}, Smith \emph{et al.} suggested the following mechanisms that can explain a factor of 10 increase in the PL' ``standard'' damping: 1) PL-FL spin-pumping and 2) strong interfacial exchange coupling at the IrMn/PL interface.

\begin{figure}[!htb]
\centering
\subcaptionbox{$\lambda_\textrm{PL1} = 0.01$\label{subfig_lambda_PL=0_01}}%
  [.49\linewidth]{\includegraphics[scale=.35]{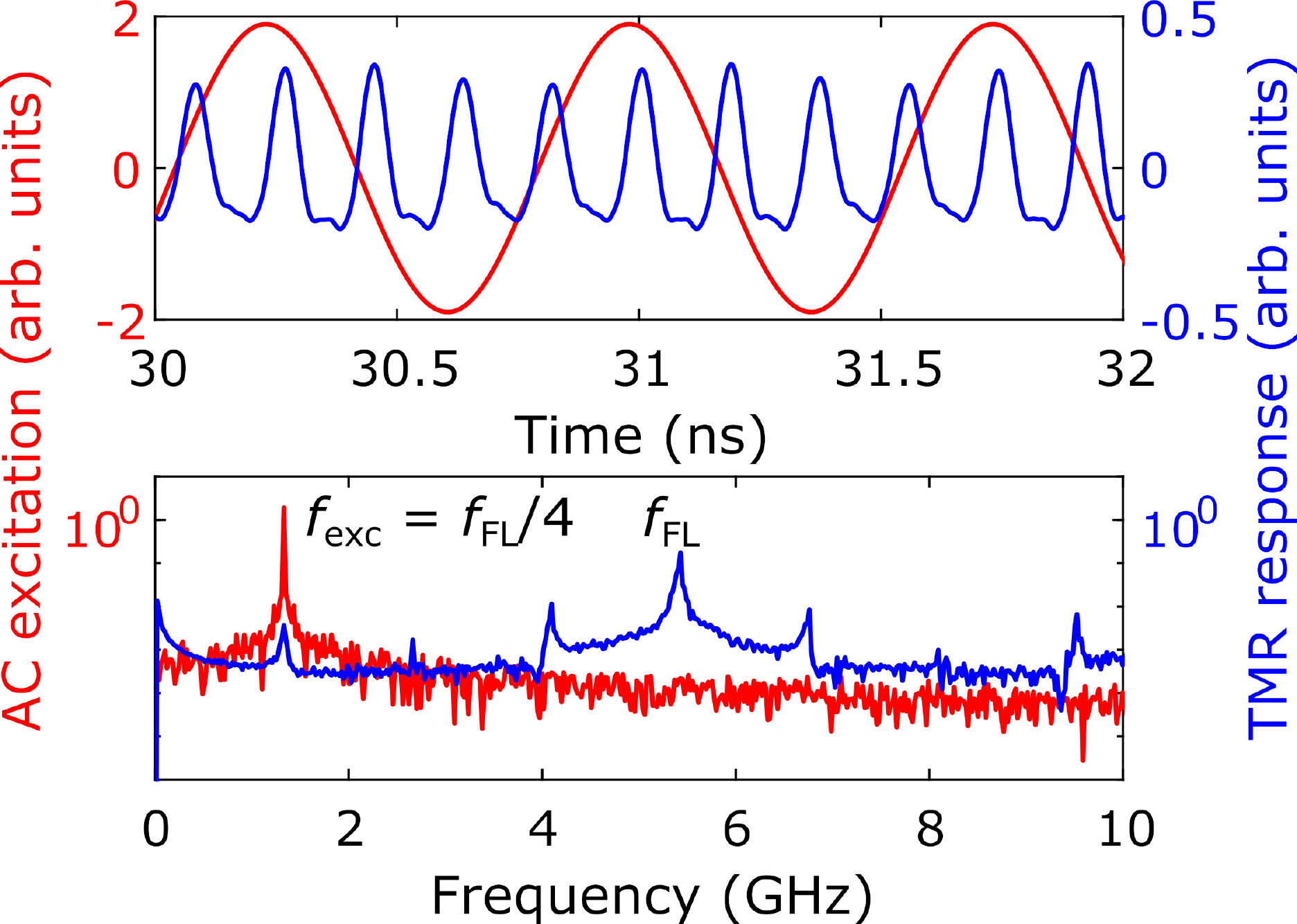}}\hfill
\subcaptionbox{$\lambda_\textrm{PL1} = 0.1$\label{subfig_lambda_PL=0_10}}
  [.49\linewidth]{\includegraphics[scale=.35]{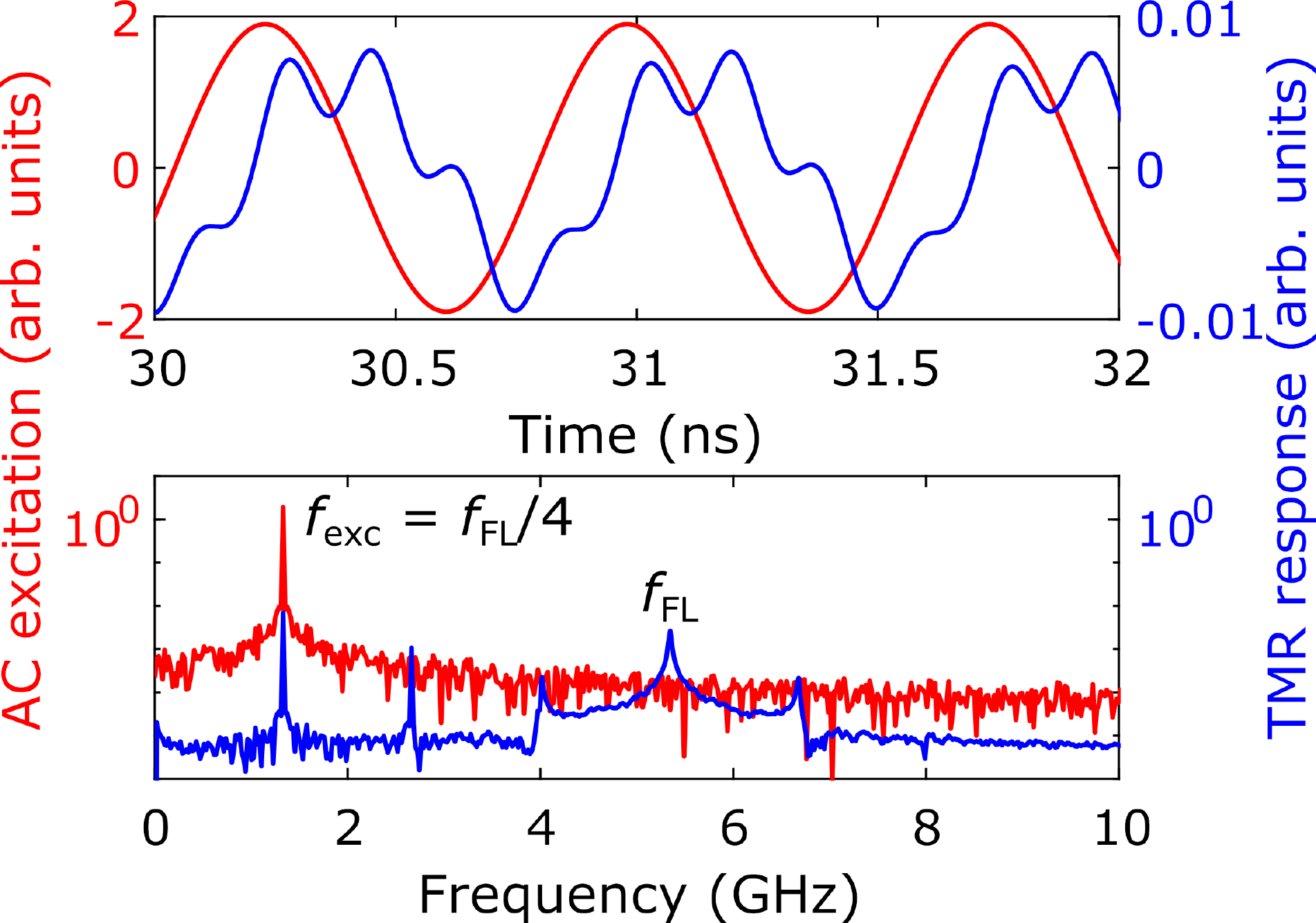}}
\caption{Time- (top) and frequency-domain (bottom) responses of the read sensor model excited at 1/4 the free layer (FL) ferromagnetic resonance (FMR) frequency $f_\textrm{FL}$. Lower pinned layer PL1 damping (\subref{subfig_lambda_PL=0_01}) evokes strongly nonlinear magnetization oscillations accompanied by much stronger harmonic response than in (\subref{subfig_lambda_PL=0_10}). In the amplitude spectra, note the ordinate axes' logarithmic scale.}
\label{fig_lambda_PL_f_exc=f_0div4}
\end{figure}

In test simulations, we have observed that the pinned layer PL1 damping strongly affects the read sensor's dynamics (Fig.~\ref{fig_lambda_PL_f_exc=f_0div4}) even though this layer is assumed to be pinned. This effect is due to the large magnetodipolar interaction not only between the PL1 and PL2, but also between the PL1 and FL. The latter interaction is strong because the FL is relatively thick ($\approx\!\!7~{\rm nm}$). These couplings lead to a significant energy transfer to the PL1 and substantial energy dissipation resulting from the overdamped dynamics of the PL1. As a consequence, neither a noticeable DC response nor a significant harmonic response is observed at $\lambda_{\rm PL1} = 0.1$ (Fig.~\ref{subfig_lambda_PL=0_10}).

Based on these observations, we compared the system studied in Ref.~\cite{smith_2010} with our sensor composition. In contrast to our MTJ-based stack, Smith \emph{et al.} studied current-perpendicular-to-plane giant-magnetoresistive (CPP-GMR) spin-valve stacks. Their samples, being just a pinned layer coupled to an AFM layer, did not have a PL2/PL1 with Ru in between. This can lead to significant differences between the effective dampings of the pinned layer adjacent to the AFM layer in these two systems. In Ref.~\cite{mohammadi_2017}, Mohammadi \emph{et. al.} reported an inverse-thickness-squared dependence of damping for exchange-biased CoFe layers and increased damping (but still lower than 0.1 for a 3-nm-thick CoFe layer) \emph{via} spin-pumping. Thus, based on the arguments presented above and partially on the results from Ref.~\cite{mohammadi_2017}, we have set the PL1 damping constant to the same value $\lambda_{\rm PL1} = 0.01$ as for other layers. Still, further studies of this question are highly desired.

\emph{Mutual ST effect between the FL and PL2.} Within a multilayered stack, the ST term  (\ref{eq_field-like_term}) can be taken into account only on the FL if and only if the following statements are true: 1) The FL is significantly thinner than all other layers. Being a surface effect, the ST is more efficient for thinner layers. 2) The PL is usually pinned to the AFM layer by an exchange bias coupling, which is much stronger than the external field onto the FL.

In our case, both statements are false. In the read sensor design, the FL is much thicker than the PL1 (7~nm \emph{versus} 3~nm). Furthermore, the external (side bias) field onto the FL is approximately the same as the exchange bias field onto the PL1, which is antiferromagnetically coupled to the PL2 (both fields are $\approx\!\!1000$~Oe). Thus, the FL-PL2 coupling must be considered in any adequate treatment of magnetization dynamics in the read head system. To account for this interaction, we have included the ST terms (\ref{eq_field-like_term}) on \emph{both} PL2 and FL. The direction of the electron polarization used to compute the ST effect on the FL has been adjusted based on the magnetization direction of the PL2, and vice versa. The ST parameters $\Lambda$ and $P$ were set to be the same for both layers.

\begin{figure}[!htb]
\centering
\subcaptionbox{\label{subfig_modelled_tot_dc_ST_effect}}%
  [.49\linewidth]{\includegraphics[scale=.35]{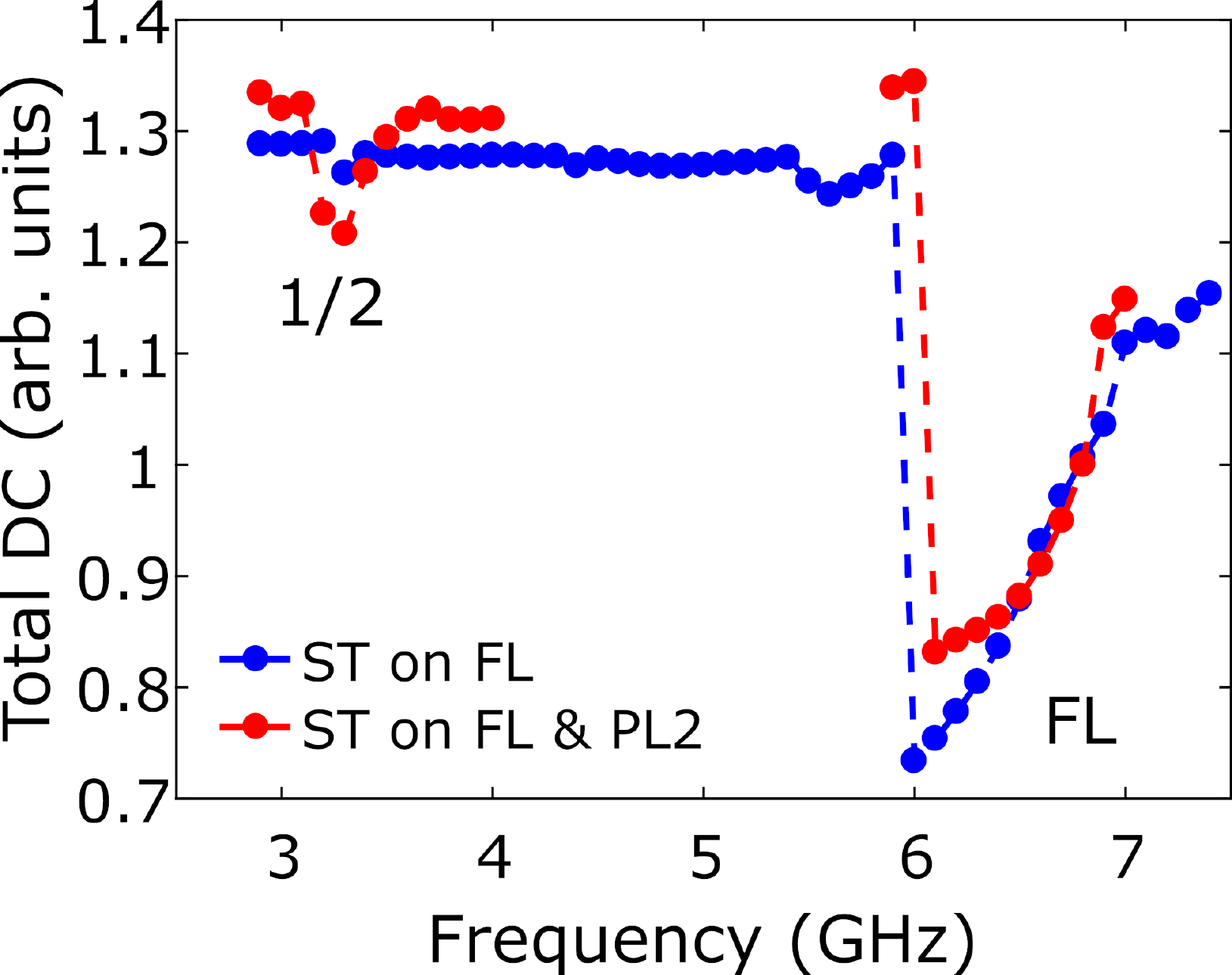}}\hfill
\subcaptionbox{\label{subfig_modelled_tot_dc_temp_effect}}
  [.49\linewidth]{\includegraphics[scale=.35]{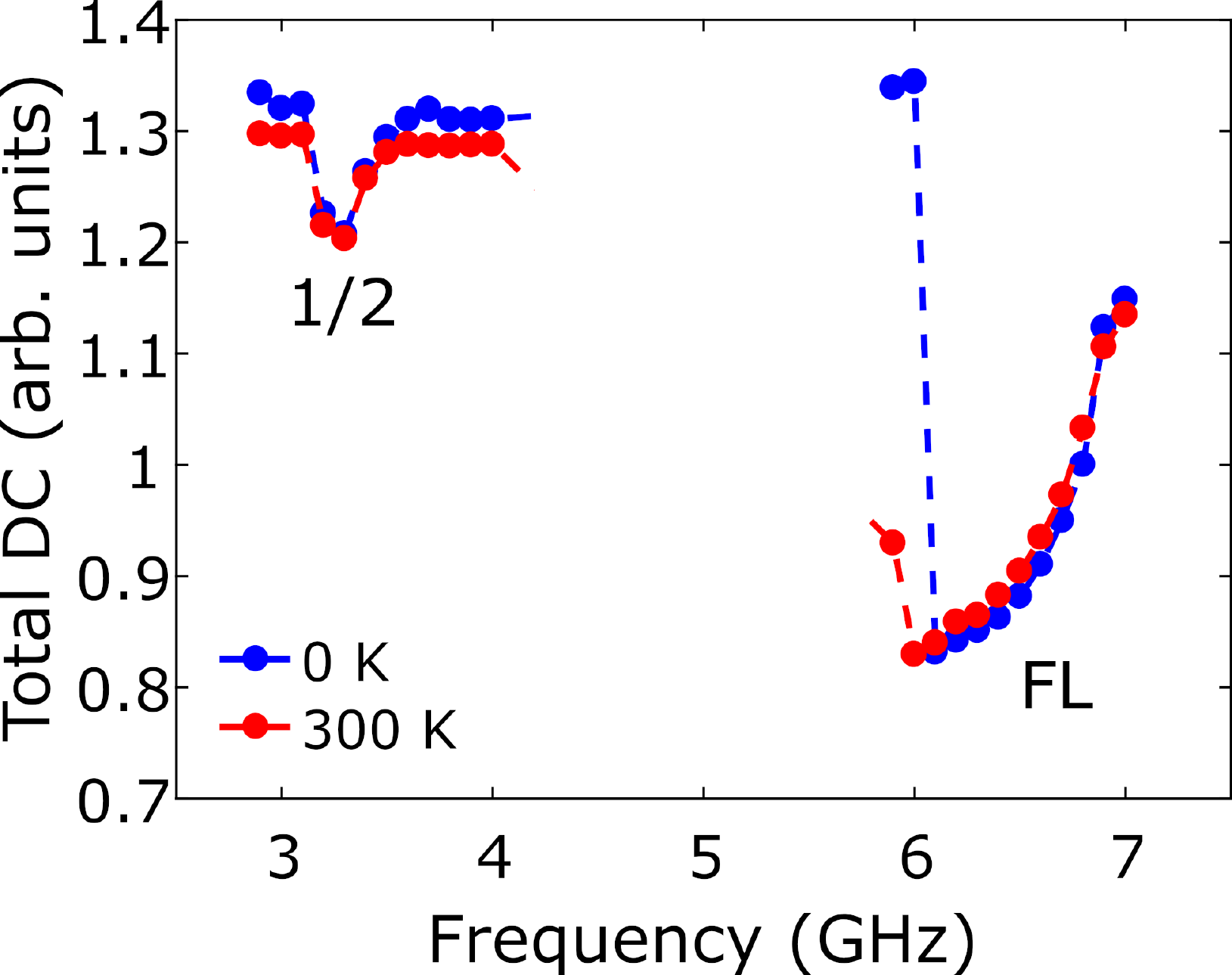}}
\caption{(\subref{subfig_modelled_tot_dc_ST_effect}) Simulated DC responses ($\Lambda = 4$, $P = 0.45$) emphasizing the importance of the spin-torque (ST) effect not only on the FL but also on the PL2. The corresponding ``constant'' and ``oscillating'' contributions to the total DC responses are shown in Appendix~\ref{sec_supp_info}, Fig.~\ref{fig_dc_response_ST_effect}. (\subref{subfig_modelled_tot_dc_temp_effect}) Simulated DC response for $\Lambda = 4$, $P = 0.45$, and different temperature conditions.}
\label{fig_modelled_tot_dc_response_miscellaneous_effects}
\end{figure}

Test simulations have confirmed the importance of the mutual ST effect between the FL and PL2 (Fig.~\ref{subfig_modelled_tot_dc_ST_effect}). In the presence of the ST effect on the PL2, the peak at $f_\textrm{FL}/2$ is $\approx\!\!5$ times larger (red dots) than when it is absent (blue dots). The influence of the mutual ST effect is especially pronounced in the the DC response's ``constant'' term (Fig.~\ref{subfig_const_dc_ST_effect} in Appendix~\ref{sec_supp_info}).

\emph{Stability of the dynamic regime with respect to thermal fluctuations.} Nonlinear dynamic systems are prone to chaotic behavior~\cite{shivamoggi_2014_intro}. In our model, \emph{e.g.}, we achieved a quasi-chaotic regime when the side bias field onto the FL was too low. Hence, it was necessary to verify the stability of the phase-locking regime with respect to thermal fluctuations. We accomplished this by comparing the simulated DC responses obtained at zero temperature and $T = 300~\textrm{K}$. Figure~\ref{subfig_modelled_tot_dc_temp_effect} shows that room-temperature fluctuations did not disturb the phase-locking regime and allowed the peak at $f_\textrm{FL}/2$ to remain intact.


\section{\label{sec_results_and_discussion}Results and discussion}


\begin{figure}[!htb]
  \centering
    \includegraphics[scale=.35]{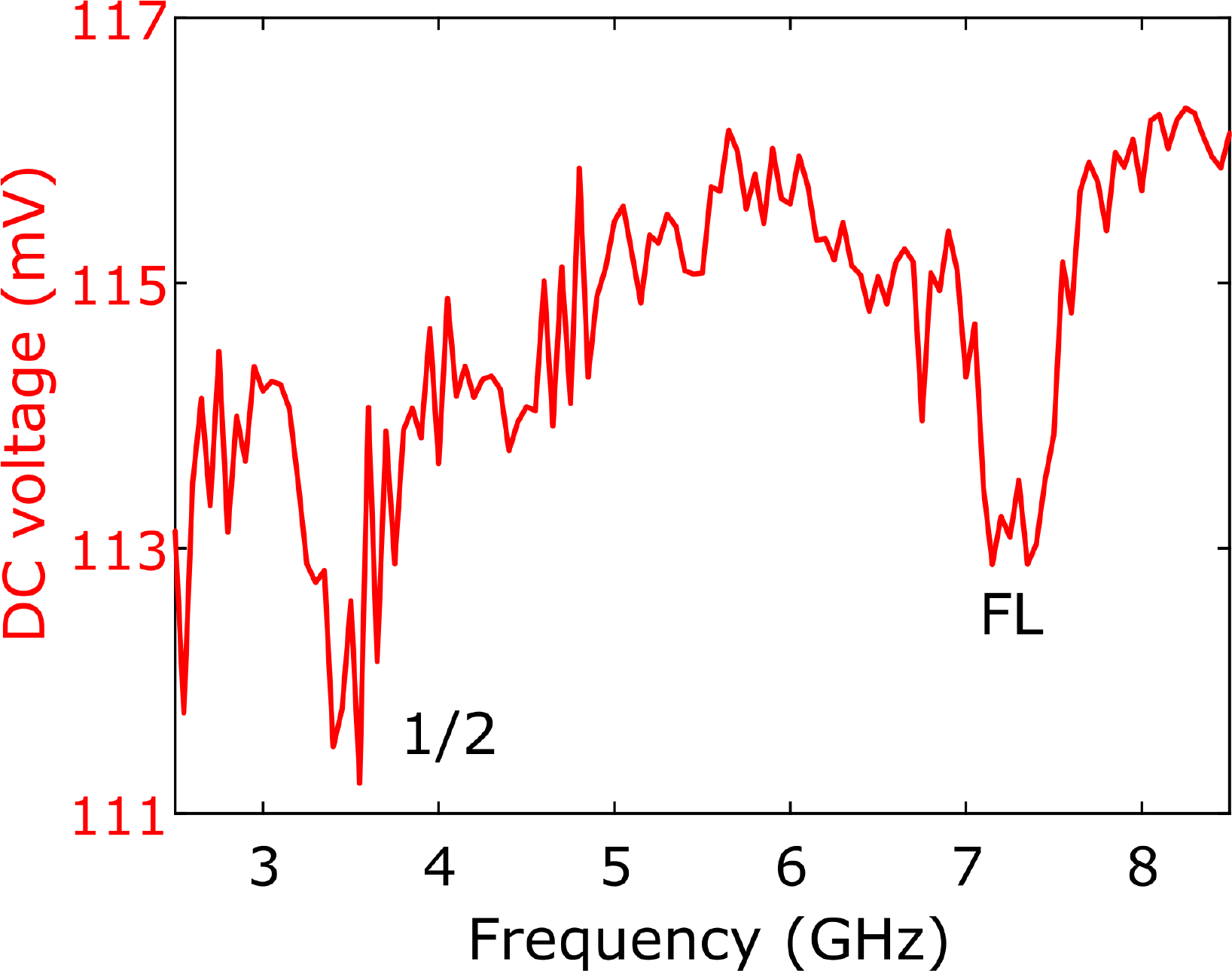}
  \caption{Experimental DC response shown within the frequency range from $f_\textrm{FL}/2$ to $f_\textrm{FL}$. Its nonlinear vector network analyzer (NVNA) measurements are presented in  Fig.~\ref{fig_harmonic_response_-5_dBm_+250_uA}.}
  \label{fig_experimental_DC_response}
\end{figure}

The micromagnetic modeling considerations proposed in Section~\ref{subsec_modeling_considerations} helped to calibrate the model. The next step is to relate the simulated DC and harmonic responses to the experimental ones.

To aid the reader, in Fig.~\ref{fig_experimental_DC_response} we show the representative sample's experimental DC response (extracted from Fig.~\ref{fig_dc_readout_-5_dBm}) limited to the frequency range from $f_\textrm{FL}/2$ to $f_\textrm{FL}$ ($f_\textrm{FL}$ being the FL's natural FMR frequency).


\subsection{\label{subsec_DC_response_results}DC response}

\emph{Effect of the ST parameters.} The spin-torque parameters appear naturally in the derivation of Slonczewski's approximation for asymmetric ferromagnetic/non-ferromagnetic/fer\-ro\-mag\-net\-ic multilayers~\cite{slonczewski_2005}. Experiments and theory have provided estimated values for both $P$ and $\Lambda$~\cite{xiao_2004,krivorotov_2007,rippard_2010}. Because of the uncertainty in both parameters, $P = 0.35$ and $\Lambda = 1.5$ are only first estimates.

\begin{figure}[!htb]
\centering
\subcaptionbox{\label{subfig_modelled_tot_dc_P_sweep}}%
  [.49\linewidth]{\includegraphics[scale=.35]{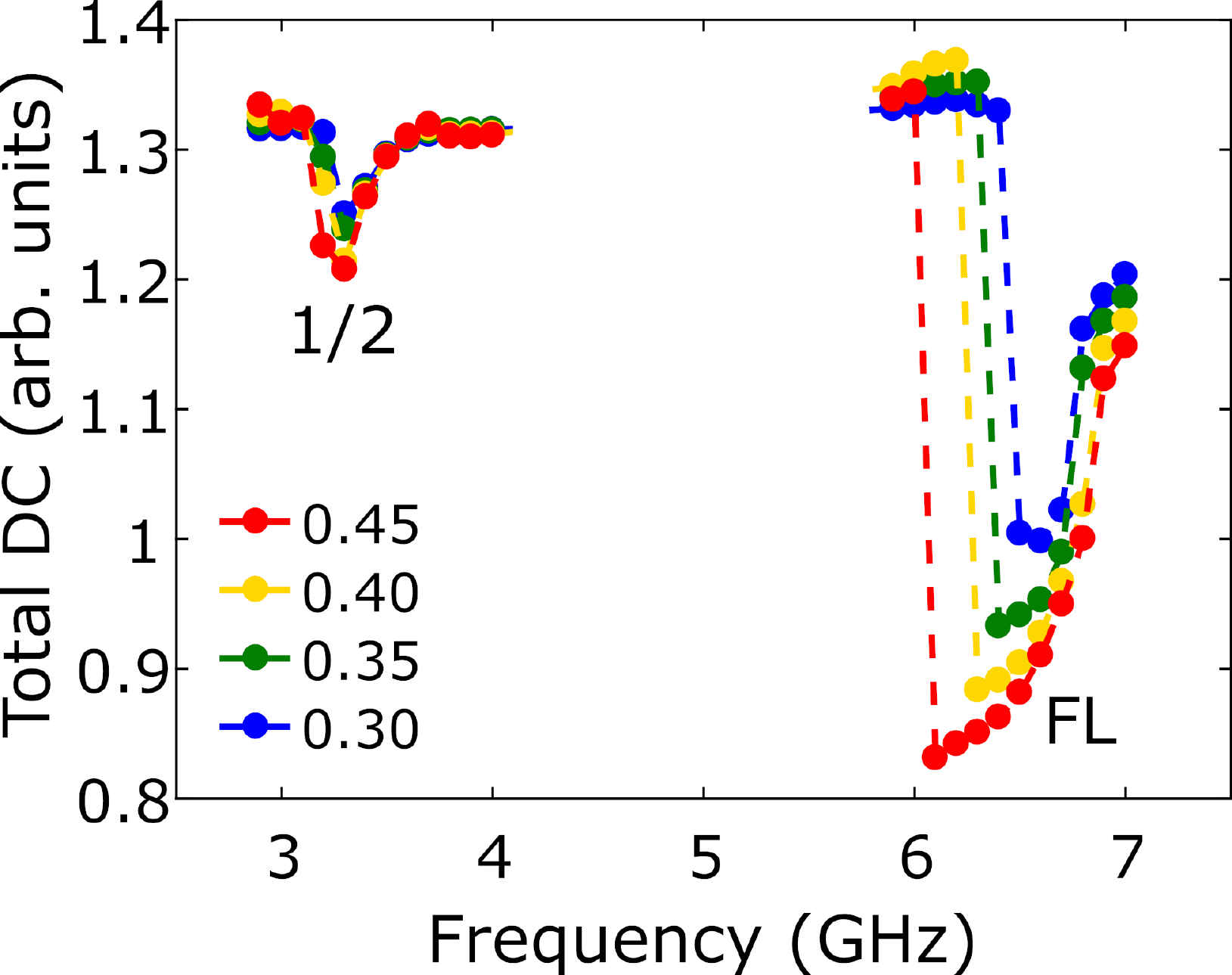}}\hfill
\subcaptionbox{\label{subfig_modelled_tot_dc_Lambda_sweep}}
  [.49\linewidth]{\includegraphics[scale=.35]{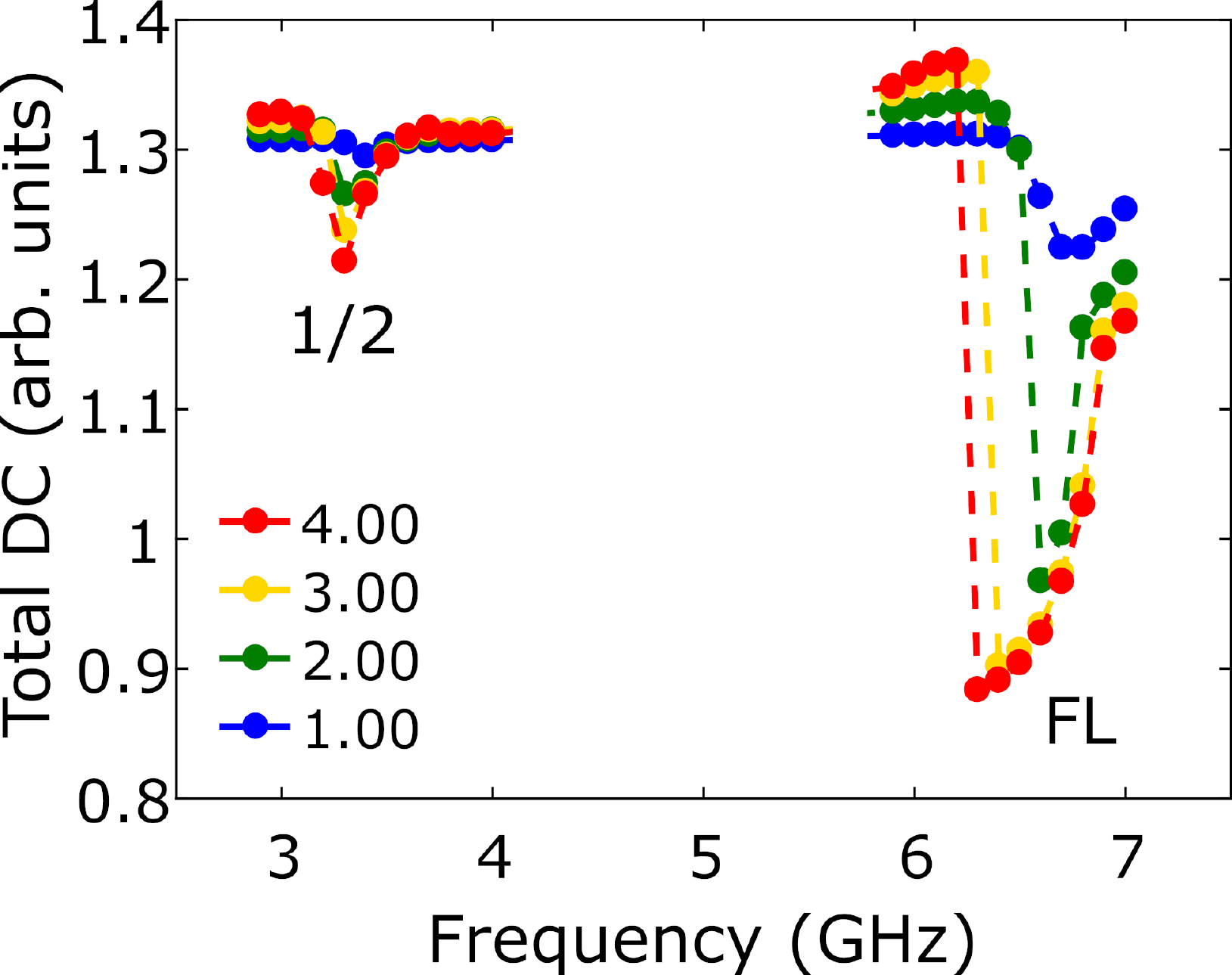}}
\caption{Simulated DC responses for different (\subref{subfig_modelled_tot_dc_P_sweep}) spin polarization factors $P$ at $\Lambda = 4$ and (\subref{subfig_modelled_tot_dc_Lambda_sweep}) asymmetry parameters $\Lambda$ at $P = 0.4$ plotted \emph{versus} the excitation frequency. The corresponding ``constant'' and ``oscillating'' contributions to the total DC responses are shown in Appendix~\ref{sec_supp_info}, Figs.~\ref{fig_dc_response_P_sweep} and~\ref{fig_dc_response_Lambda_sweep}).}
\label{fig_modelled_tot_dc_response}
\end{figure}

The uncertainty in the ST parameters provided the main degree of freedom in achieving sufficient qualitative agreement between the experimental and simulated DC responses. The characteristic ``foldover'' FL FMR profiles evolve with increasing spin polarization factors $P$ and asymmetry parameters $\Lambda$ (Fig.~\ref{fig_modelled_tot_dc_response}). Moreover, higher $P$ and $\Lambda$ significantly broaden the FL mode's linewidth. We estimated our samples' ST parameters by matching the FL mode's linewidths in the experimental and simulated DC responses. The ${\sim}0.5~\textrm{GHz}$ experimental linewidth corresponds to $P = 0.4$ and $\Lambda = 4$. These values are higher than our first estimates provided in the published works: $P = 0.35$ and $\Lambda = 1.5$. Such an increase, however, is necessary to match the experimental FL mode's linewidth along with achieving the measurable DC response at $f_\textrm{FL}/2$.

When excited around a fractional frequency of the FL mode, increasing ST parameters facilitate the strongly nonlinear magnetization dynamics accompanied by the generation of the higher-order harmonics and measurable DC (\emph{e.g.}, as in Fig.~\ref{subfig_lambda_PL=0_01}). The two contributions to the total DC response contain additional physical insights. Figures~\ref{fig_dc_response_P_sweep} and~\ref{fig_dc_response_Lambda_sweep} in Appendix~\ref{sec_supp_info} show that the DC response at the FL FMR frequency is primarily determined by mixing of the TMR oscillations with the excitation signal [``oscillating" term in Eq.~(\ref{eq_dc_voltage_response})]. On the other hand, the DC response at 1/2 the frequency of the FL FMR mode is defined by the ``constant" term.

\emph{Response at the FL mode's ``sub-harmonics.''} 
The magnetodipolar interaction between the FL and PL2 causes positive feedback, which qualitatively affects the system dynamics. This feedback can be thought of as follows. The deviation of the FL magnetization from its preferred orientation generates a stray field. This field causes the PL2 magnetization to deviate in the direction opposite to that of the FL. This, in turn, results in an even larger deviation of the FL magnetization due to the influence of the PL2 stray field, thus producing positive feedback between the magnetization dynamics of the FL and PL2~\cite{berkov_2009}

\begin{figure}[!htb]
  \centering
    \includegraphics[scale=.7]{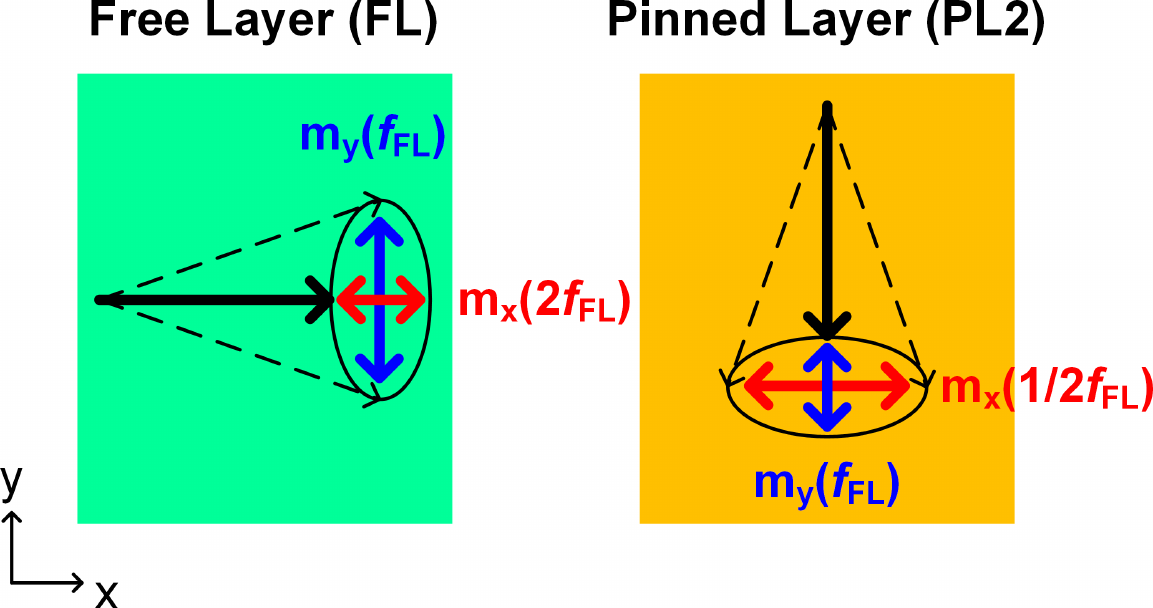}
  \caption{Magnetization precession's elliptical trajectory and magnetodipolar feedback between the FL and PL2 facilitate the response at 1/2 the FL frequency.}
  \label{fig_sub-harmonic_generation}
\end{figure}

The FL precession exhibits an elliptical trajectory. Thus, the $m_x$ component moves back and forth twice during one oscillation cycle, hence its oscillation frequency is approximately twice that of the $m_y$ component~\cite{berkov_2005}. Thus, if $f_\textrm{FL}$ is the FL's resonant precession frequency, then $f(m_x) = 2f(m_y) = 2 f_\textrm{FL}$. In this physical picture, the magnetodipolar interaction induces coupling between the FL and PL2. Due to the large $y$-component of the FL's stray field induced by strong $m_y$ oscillations with the frequency $f_\textrm{FL}$, the above-mentioned coupling induces the $m_y$ oscillations of the PL2 with the same frequency $f_\textrm{FL}$. These oscillations, in turn, result in the $m_x$ oscillations of the PL2 with the frequency $f_\textrm{FL}/2$ (Fig.~\ref{fig_sub-harmonic_generation}). The feedback \emph{via} the PL2's magnetodipolar field leads to the oscillation of the FL with the frequency $f_\textrm{FL}/2$ as well as all its harmonics. Thus, the FL power spectrum would contain the FL mode, its ``sub-harmonic" at $f_\textrm{FL}/2$, and their higher-order modes, all of which are produced due to the nonlinear nature of the FL's $m_y$ oscillations and the contribution of the PL2's $m_x$ and $m_y$ oscillations excited by the presence of the magnetodipolar interaction.

Therefore, to facilitate the DC response at 1/2, 1/3, 1/4, and 1/6 the frequency of the FL FMR mode, it is enough to combine the magnetodipolar feedback between the FL and PL2 with a low-order nonlinearity. The DC response at 1/5 the frequency of the FL FMR mode is not present in the measurement: it requires a fifth-order nonlinearity, but the fifth harmonic is usually too small to evoke phase locking.

\subsection{Harmonic response}

\begin{figure}[!htb]
\centering
\subcaptionbox{fundamental\label{subfig_fund_P=0_40_Lambda=4}}%
  [.49\linewidth]{\includegraphics[scale=.35]{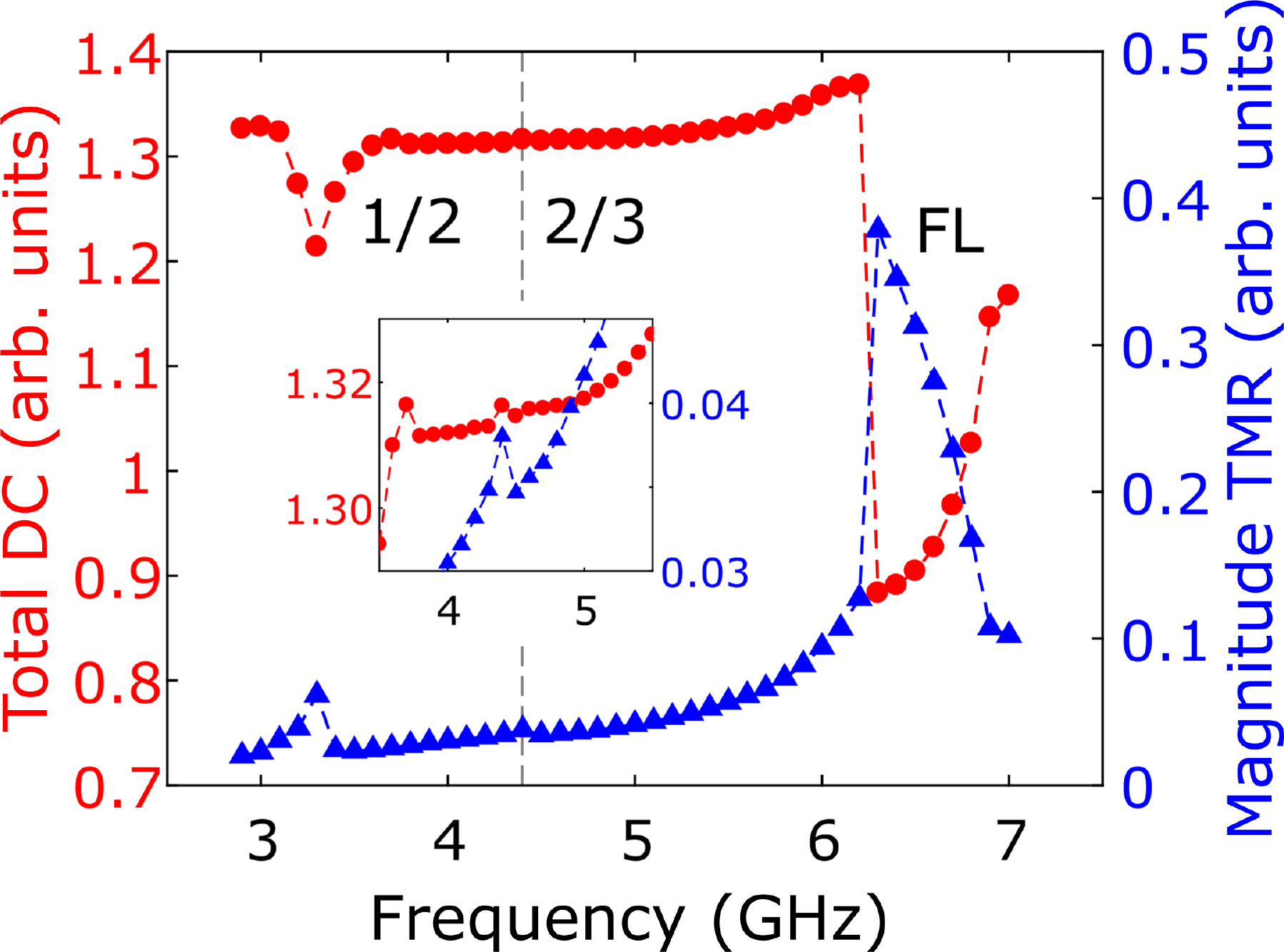}}\hfill
\subcaptionbox{second harmonic\label{subfig_2nd_harm_P=0_40_Lambda=4}}
  [.49\linewidth]{\includegraphics[scale=.35]{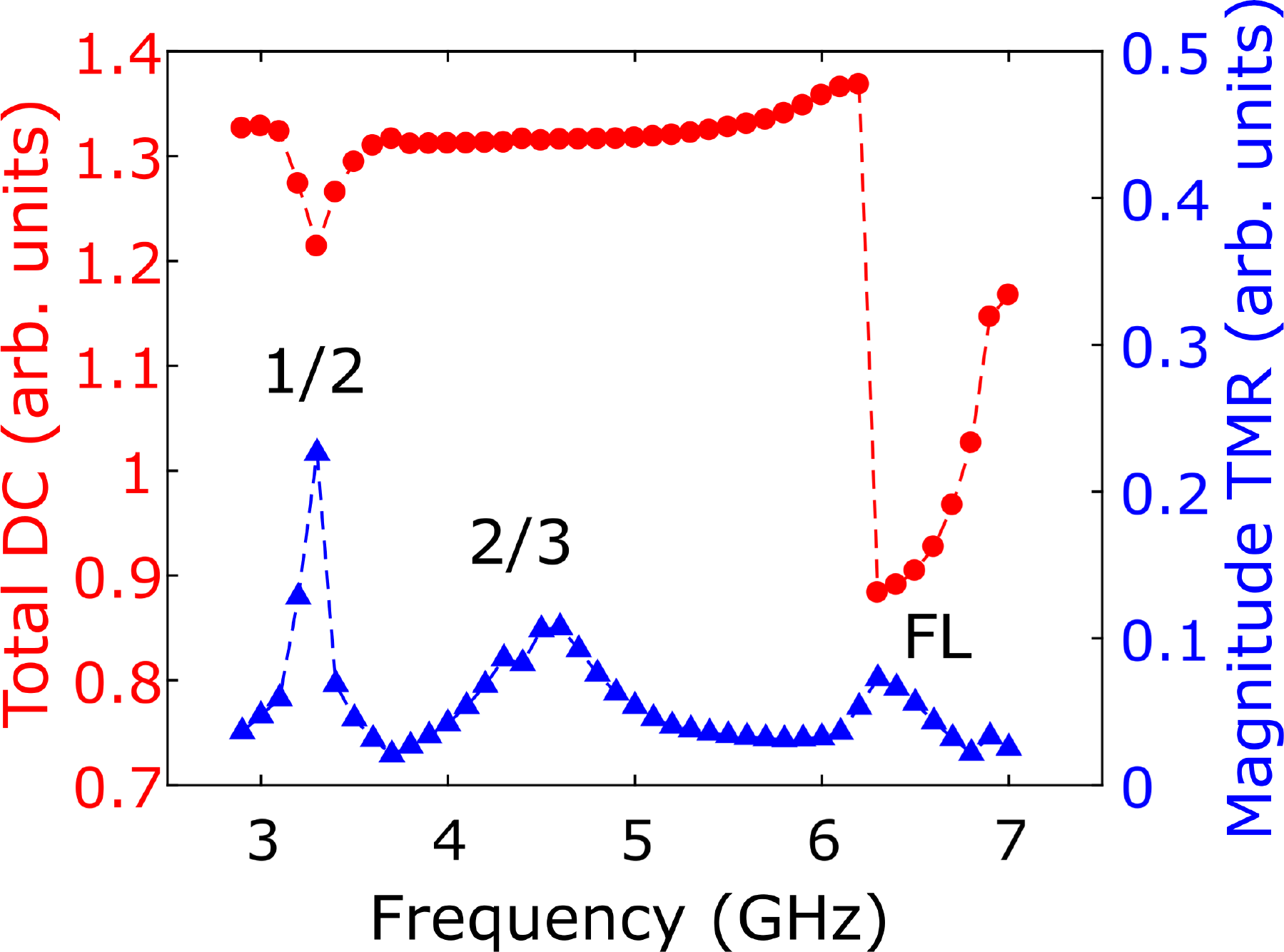}}
\caption{Simulated tunnel magnetoresistance (TMR) response at the (\subref{subfig_fund_P=0_40_Lambda=4}) excitation frequency (fundamental) and (\subref{subfig_2nd_harm_P=0_40_Lambda=4}) its second harmonic plotted \emph{versus} the excitation frequency for $\Lambda = 4$ and $P = 0.4$. The presence of peaks corresponding to 2/3 the free layer (FL) ferromagnetic resonance (FMR) frequency $f_\textrm{FL}$ suggests that the resonant condition might occur at excitation frequencies other than $f_\textrm{FL}/n$, where $n$ is a positive integer.}
\label{fig_harm_response_P=0_40_Lambda=4}
\end{figure}

Figure~\ref{fig_harm_response_P=0_40_Lambda=4} shows the fundamental and second harmonic of TMR counterposed to the total DC response. Whereas the TMR response at 1/2 the FL FMR frequency $f_\textrm{FL}$ has a weak fundamental and a strong second harmonic, the TMR response at the FL FMR frequency has a strong fundamental and a weak second harmonic.

The former observation implies that the MTJ's nonlinear characteristics give rise to nonlinear oscillations under the AC excitation signal with the frequency corresponding to 1/2 the FL natural precession frequency. As the TMR response at the second harmonic is stronger than at the fundamental, the FL FMR mode seems to phase lock to this harmonic. As for the latter observation, the FL FMR mode phase locks to the TMR oscillations at the excitation frequency, thus producing a strong fundamental at the FL FMR frequency.

The major discrepancy between the measurements in Figs.~\ref{fig_1st_harm_-5_dBm_+250_uA} and~\ref{fig_2nd_harm_-5_dBm_+250_uA} and simulations in Fig.~\ref{fig_harm_response_P=0_40_Lambda=4} is the absence of the strong fundamental of $B_1$ at the FL FMR frequency $f_\textrm{FL}$. We attribute this difference to weak coupling (due to convergence to steady-state phase locking) between the NVNA's incident wave and magnetization precession. We suggest that the experimental magnetization precession has a less stable phase relation with the NVNA's incident signal than in the model.

In Fig.~\ref{fig_1st_harm_-5_dBm_+250_uA}, the ripply response in the fundamental of $B_1$ suggests that there are more peaks at fractional frequencies of the FL mode than what we have identified. In particular, the enlarged DC responses in Fig.~\ref{fig_experimental_DC_response} indicate a resonant feature between 1/2 the FL FMR mode and the FL FMR mode. We qualitatively replicated this feature in the simulated TMR response [Figs.~\ref{subfig_fund_P=0_40_Lambda=4} (inset) and \ref{subfig_2nd_harm_P=0_40_Lambda=4}]. Contrary to the experiment, however, the FL FMR mode seems to phase lock to the second harmonic, not to the TMR oscillations at the excitation frequency corresponding to 2/3 the FL FMR frequency $f_\textrm{FL}$. This is clearly seen in the time- and frequency-domain TMR responses for two selected excitation frequencies: 2/3 the frequency of the FL FMR mode and 5~GHz (Fig.~\ref{fig_f_exc=2_f0_div_3} in Appendix~\ref{sec_supp_info}). The magnetic system's harmonic response (specifically, of the second harmonic) is stronger if the excitation frequency corresponds to a fractional frequency of the FL FMR mode.

\section{Conclusion}

We discovered that all characterized magnetic sensors' DC responses reveal peaks at frequencies that are the integer fractions (1/2, 1/3, 1/4, and 1/6) of the devices' natural FL FMR frequency $f_\textrm{FL}$. These peaks, in turn, generate the corresponding second and third harmonics of $B_1$. To understand the underlying physics that enabled the DC response at ``sub-harmonics'' of the FL mode, we employed micromagnetic modeling.

A comprehensive micromagnetic study suggested that the experimentally observed DC response at 1/2, 1/3, 1/4, and 1/6 of $f_\textrm{FL}$ can be defined by a low-order nonlinearity and strong magnetodipolar feedback between the FL and PL. As the PL is significantly thinner than the FL, additionally accounting for the ST effect on this layer notably enhanced the ST-driven harmonic response. Interestingly, the orthogonality of the FL and PL also facilitated the magnetic sensors' response at 1/2 the FL FMR frequency $f_\textrm{FL}$.

Most importantly, strong magnetodipolar feedback permitted sub-harmonic injection locking within a wide range of integer fractions, which can be used in the development of a new generation of frequency multipliers.

\clearpage

\appendix
\section{\label{sec_supp_info}}

\begin{figure}[!htb]
\centering
\subcaptionbox{\label{subfig_const_dc_P_sweep}}%
  [.33\linewidth]{\includegraphics[scale=.3]{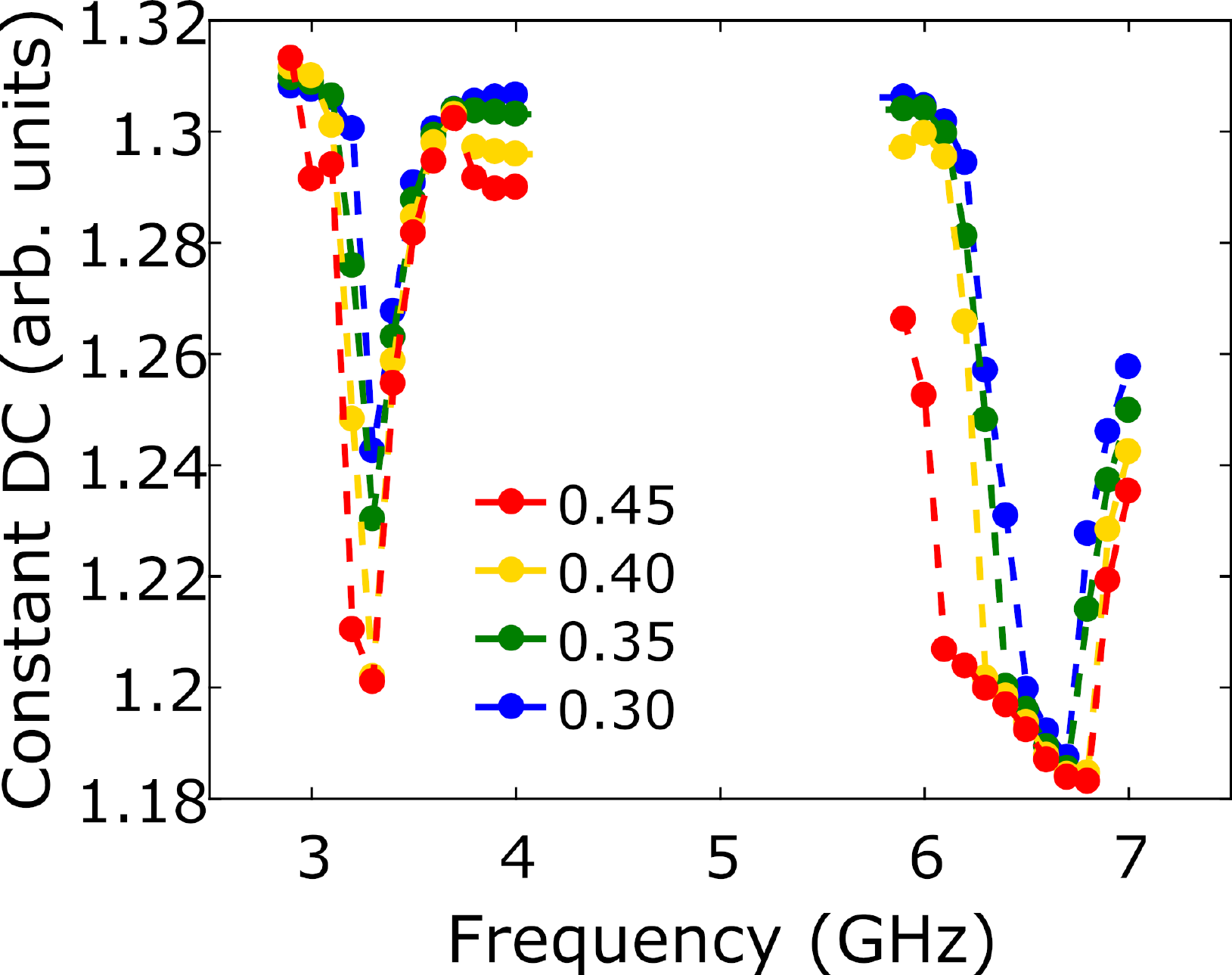}}\hfill
\subcaptionbox{\label{subfig_osc_dc_P_sweep}}%
  [.33\linewidth]{\includegraphics[scale=.3]{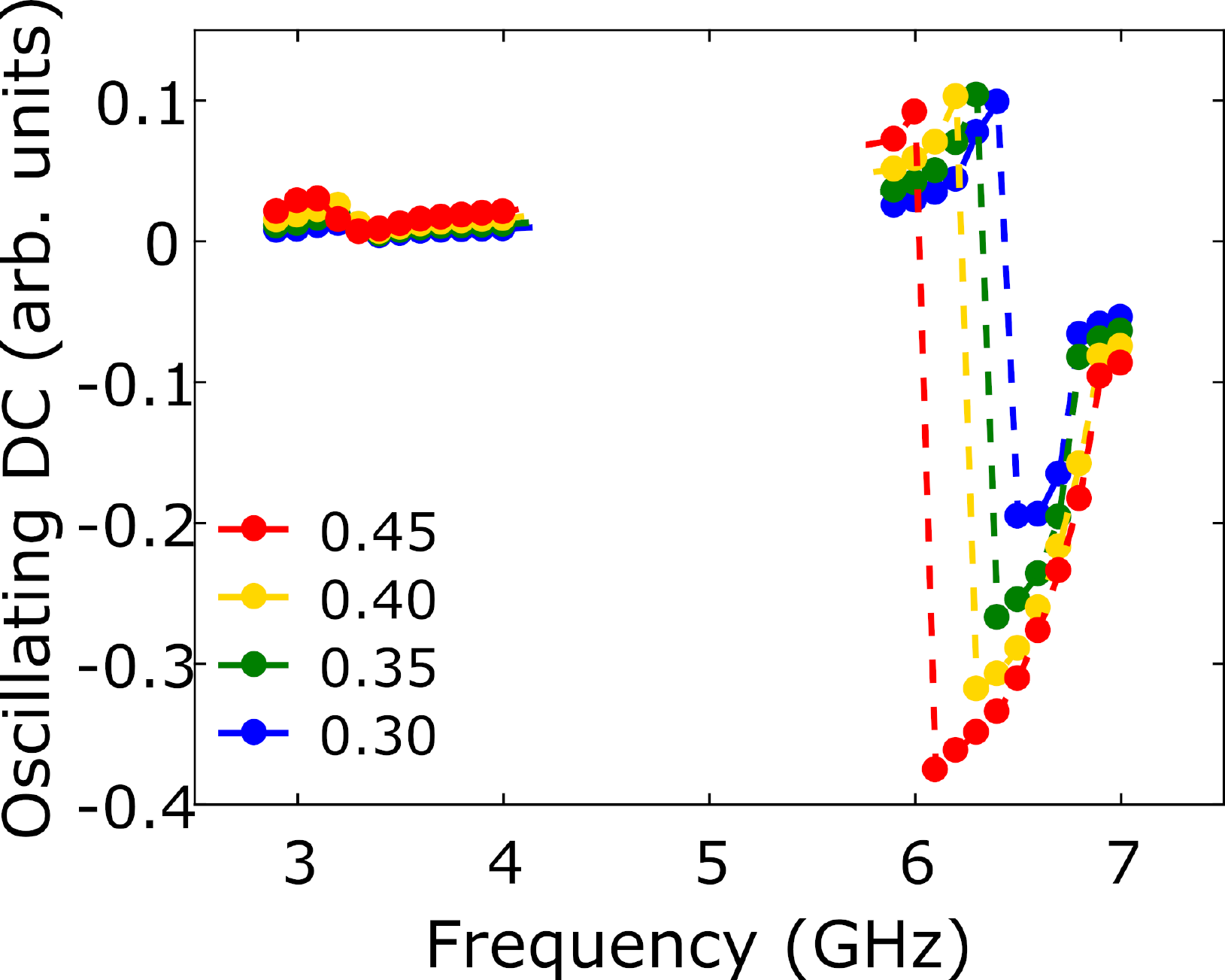}}\hfill
\subcaptionbox{\label{subfig_tot_dc_P_sweep}}
  [.33\linewidth]{\includegraphics[scale=.3]{my_figs/tot_dc_Lambda=4_P_sweep_rev.pdf}}
\caption{Simulated (\subref{subfig_const_dc_P_sweep}) ``constant" and (\subref{subfig_osc_dc_P_sweep}) ``oscillating" contributions to the (\subref{subfig_tot_dc_P_sweep}) total DC response for different spin polarization factors $P$ at $\Lambda = 4$}
\label{fig_dc_response_P_sweep}
\end{figure}

\begin{figure}[!htb]
\centering
\subcaptionbox{\label{subfig_const_dc_Lambda_sweep}}%
  [.33\linewidth]{\includegraphics[scale=.3]{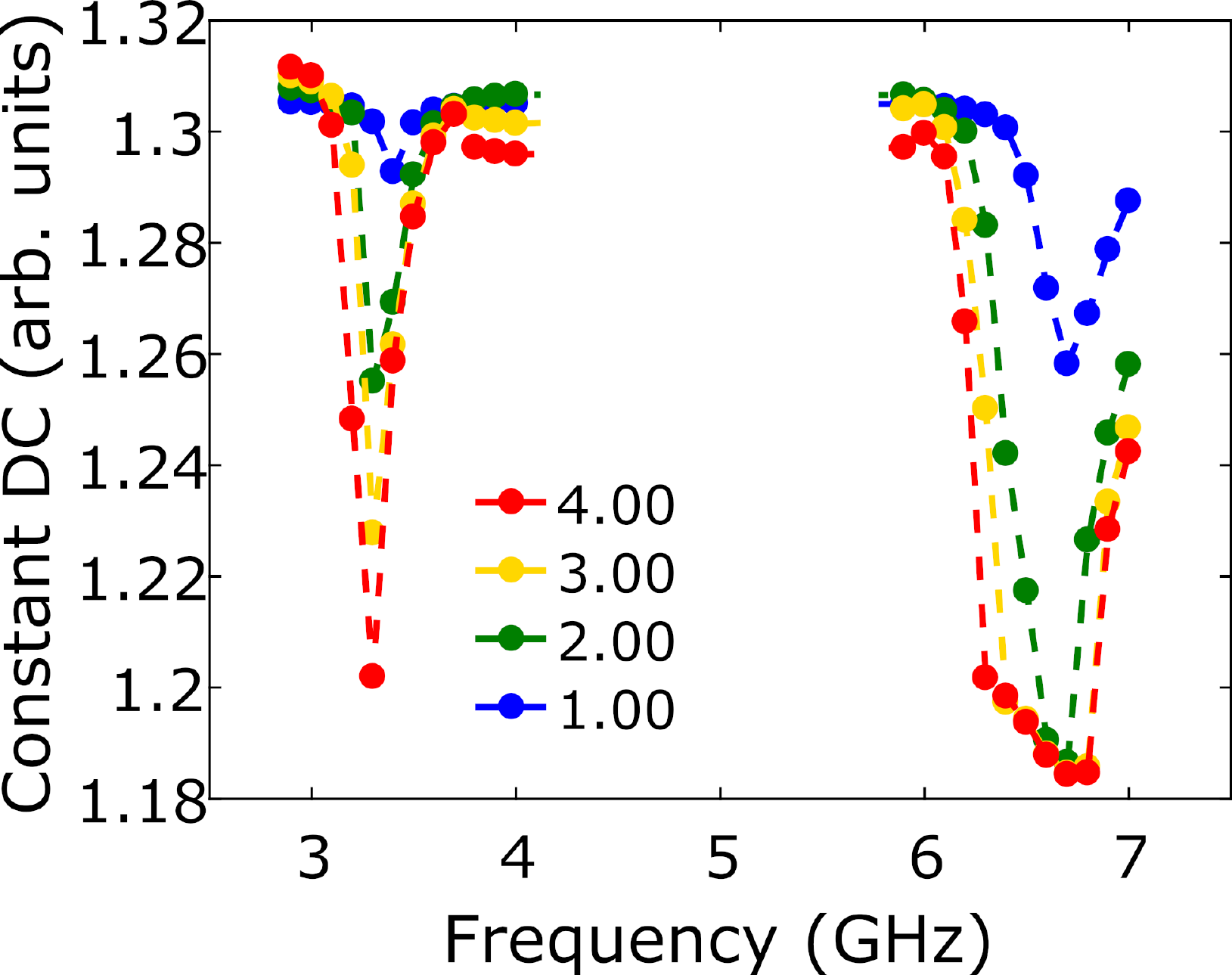}}\hfill
\subcaptionbox{\label{subfig_osc_dc_Lambda_sweep}}%
  [.33\linewidth]{\includegraphics[scale=.3]{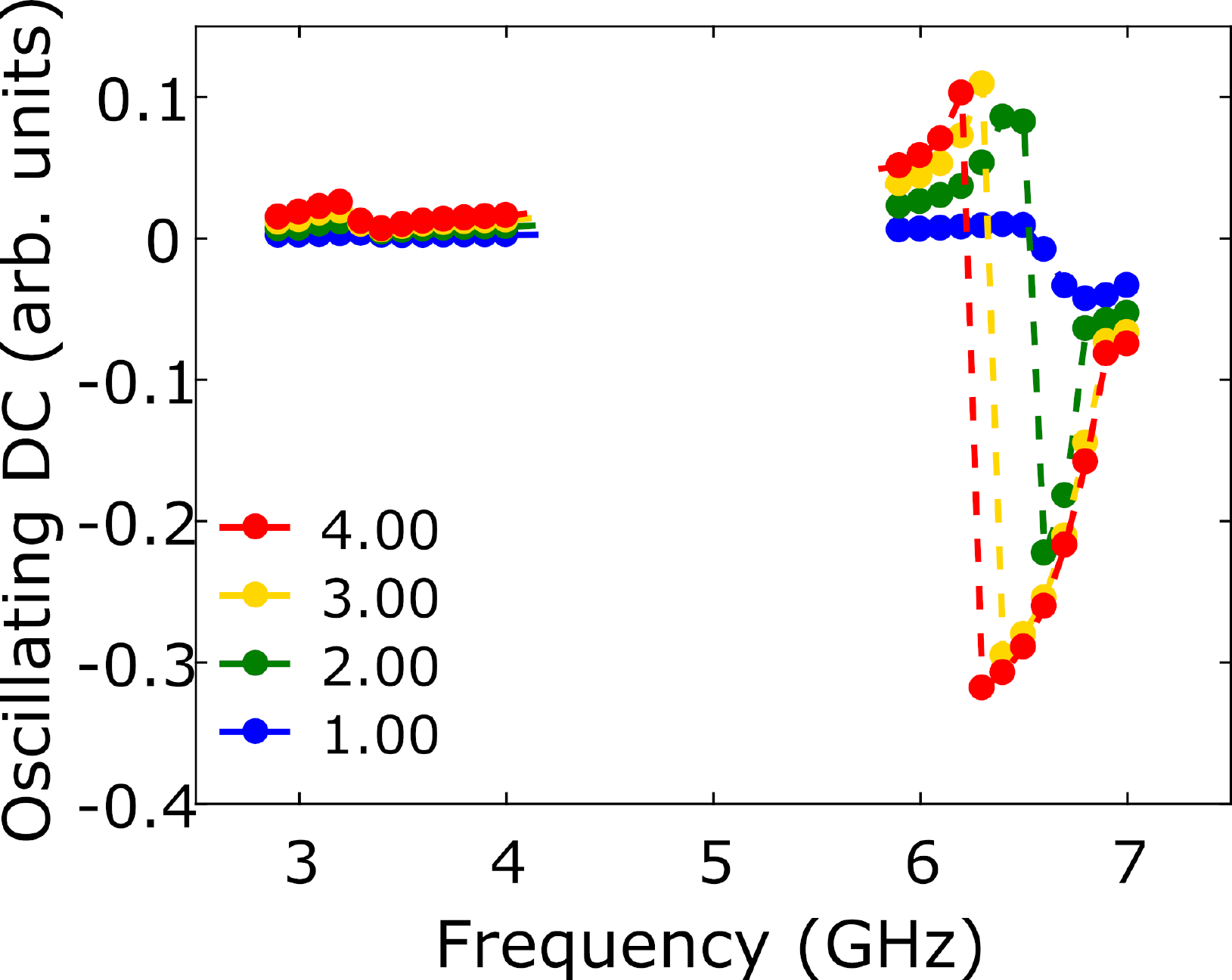}}\hfill
\subcaptionbox{\label{subfig_tot_dc_Lambda_sweep}}
  [.33\linewidth]{\includegraphics[scale=.3]{my_figs/tot_dc_P=0_40_Lambda_sweep_rev.pdf}}
\caption{Simulated (\subref{subfig_const_dc_Lambda_sweep}) ``constant" and (\subref{subfig_osc_dc_Lambda_sweep}) ``oscillating" contributions to the (\subref{subfig_tot_dc_Lambda_sweep}) total DC response for different asymmetry parameters $\Lambda$ at $P = 0.4$}
\label{fig_dc_response_Lambda_sweep}
\end{figure}

\begin{figure}[!htb]
\centering
\subcaptionbox{\label{subfig_const_dc_ST_effect}}%
  [.33\linewidth]{\includegraphics[scale=.3]{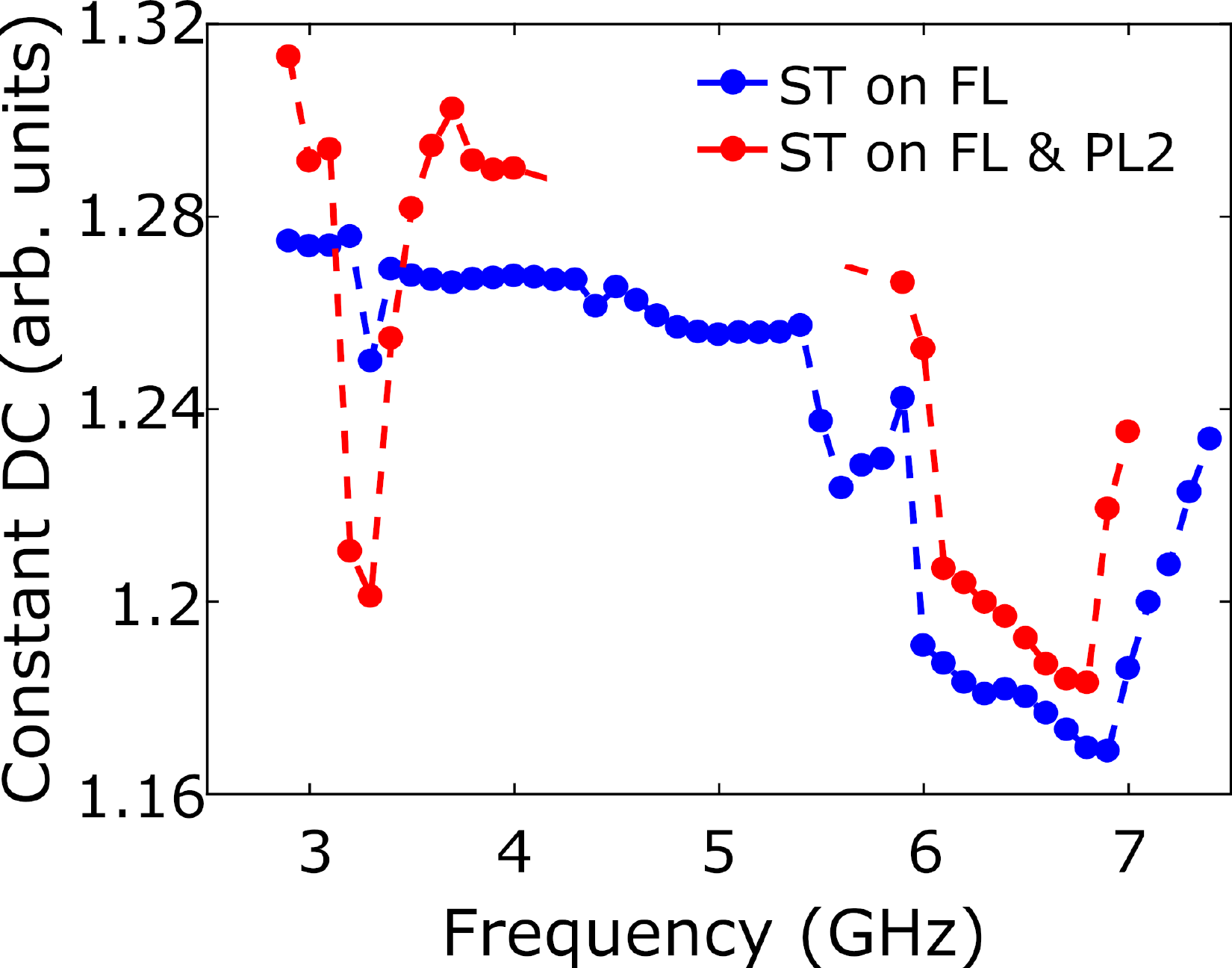}}\hfill
\subcaptionbox{\label{subfig_osc_dc_ST_effect}}%
  [.33\linewidth]{\includegraphics[scale=.3]{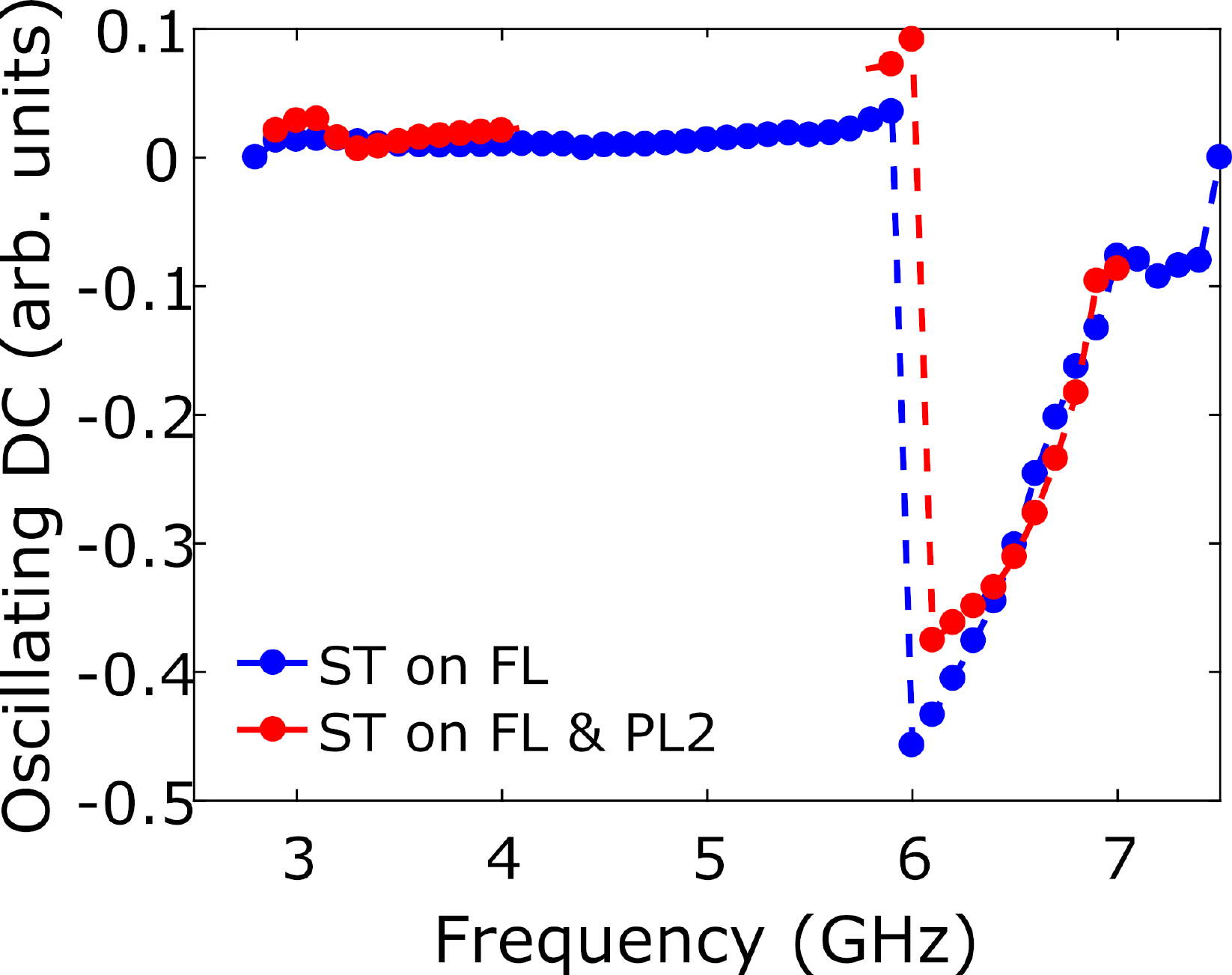}}\hfill
\subcaptionbox{\label{subfig_tot_dc_ST_effect}}
  [.33\linewidth]{\includegraphics[scale=.3]{my_figs/tot_dc_vs_ST_onto_FL_only_rev1.pdf}}
\caption{Simulated (\subref{subfig_const_dc_ST_effect}) ``constant" and (\subref{subfig_osc_dc_ST_effect}) ``oscillating" contributions to the (\subref{subfig_tot_dc_ST_effect}) total DC response for $\Lambda = 4$, $P = 0.45$, and different spin-torque (ST) scenarios}
\label{fig_dc_response_ST_effect}
\end{figure}






\begin{figure}[!htb]
\centering
\subcaptionbox{\label{subfig_f_exc=4_4_GHz}}%
  [.49\linewidth]{\includegraphics[scale=.35]{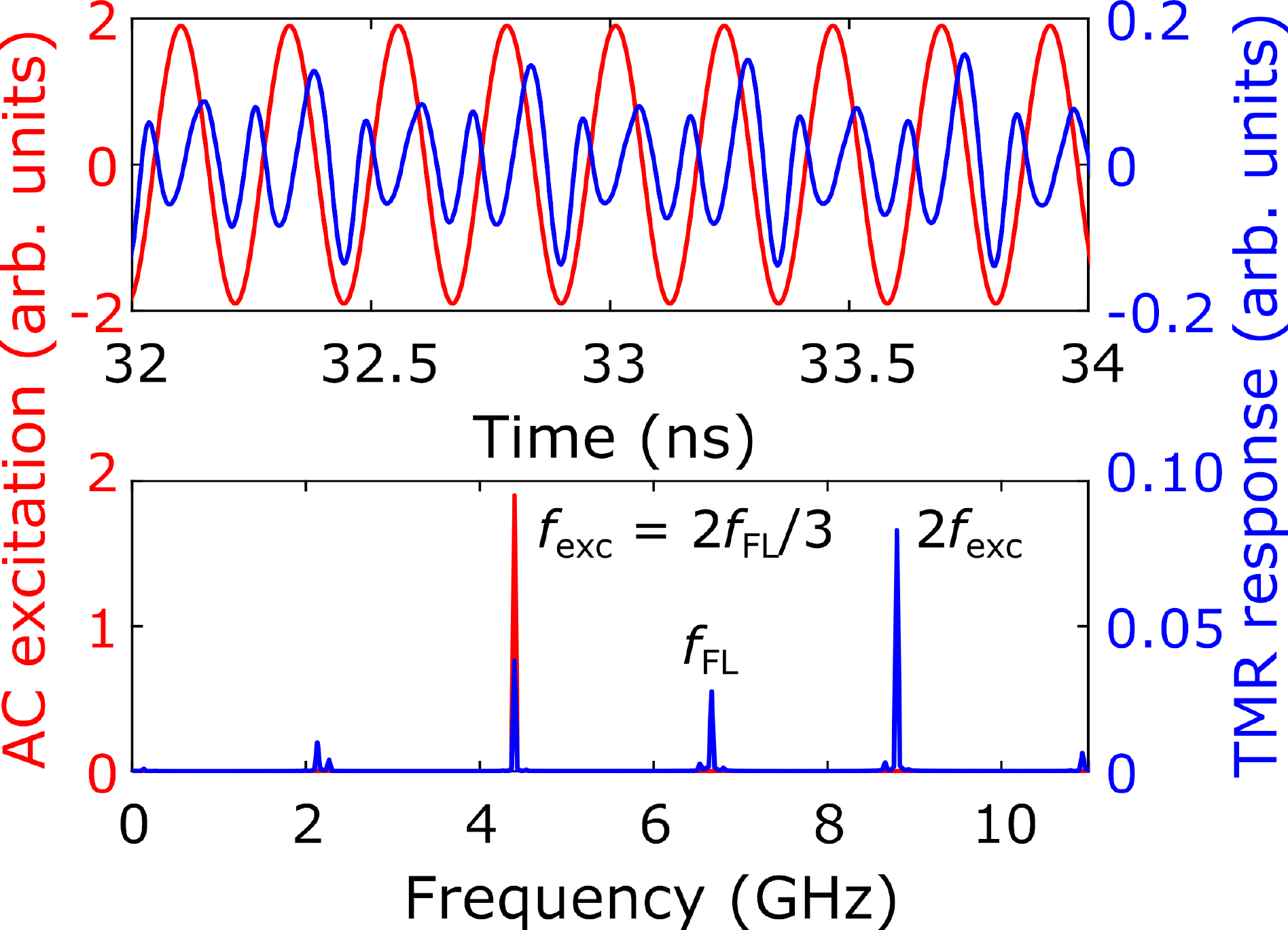}}\hfill
\subcaptionbox{\label{subfig_f_exc=5_0_GHz}}
  [.49\linewidth]{\includegraphics[scale=.35]{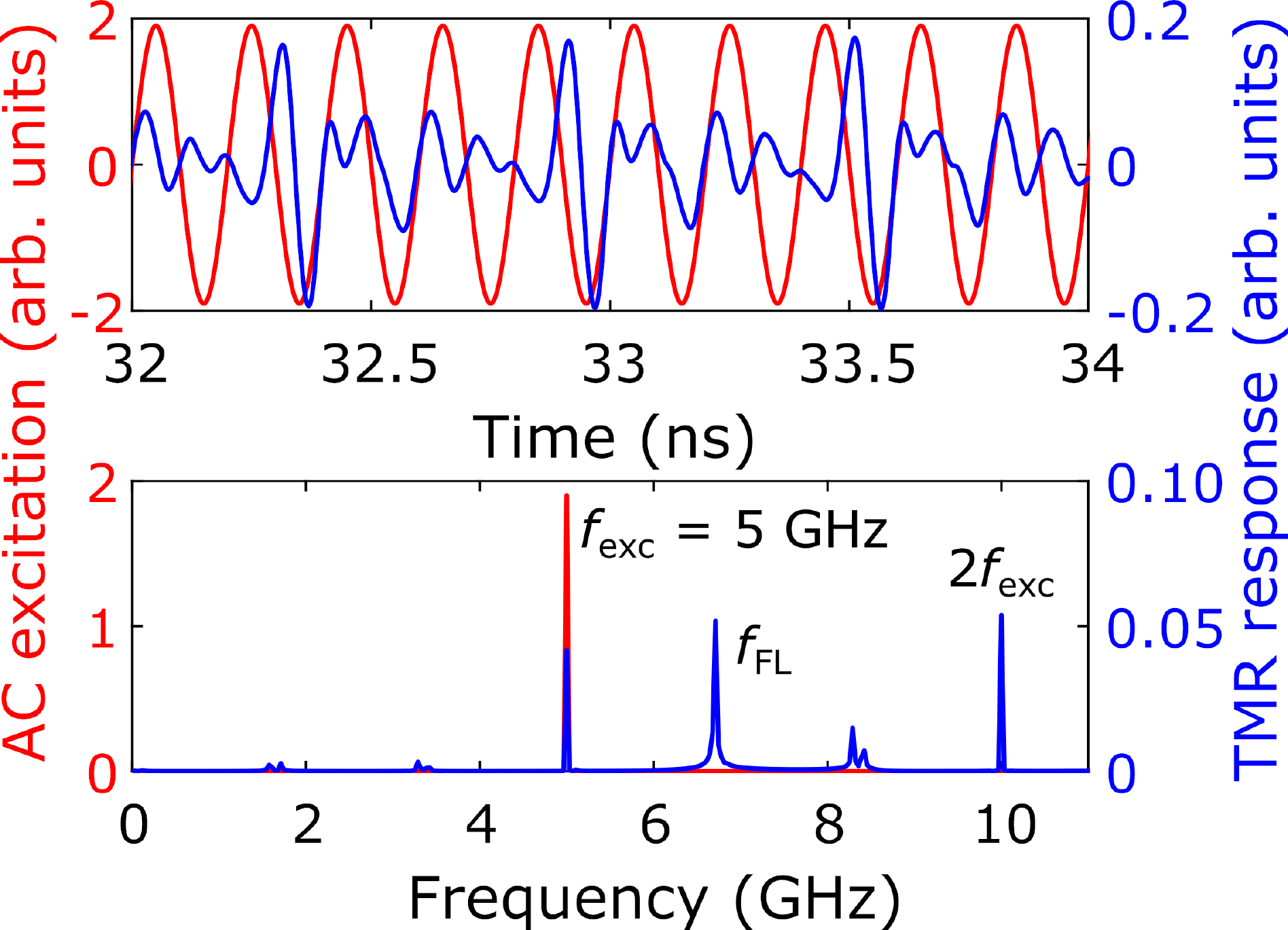}}
\caption{Time- (top) and frequency-domain (bottom) responses of the read sensor model excited at (\subref{subfig_f_exc=4_4_GHz}) 2/3 the free layer (FL) ferromagnetic resonance (FMR) frequency $f_\textrm{FL}$ and (\subref{subfig_f_exc=5_0_GHz}) 5~GHz. The magnetic system's harmonic response (specifically, of the second harmonic) is stronger if the excitation frequency corresponds to a fractional frequency of the FL FMR mode.}
\label{fig_f_exc=2_f0_div_3}
\end{figure}

\clearpage

\bibliography{my_aps}

\providecommand{\noopsort}[1]{}\providecommand{\singleletter}[1]{#1}%
\begin{thebibliography}{40}%
\makeatletter
\providecommand \@ifxundefined [1]{%
 \@ifx{#1\undefined}
}%
\providecommand \@ifnum [1]{%
 \ifnum #1\expandafter \@firstoftwo
 \else \expandafter \@secondoftwo
 \fi
}%
\providecommand \@ifx [1]{%
 \ifx #1\expandafter \@firstoftwo
 \else \expandafter \@secondoftwo
 \fi
}%
\providecommand \natexlab [1]{#1}%
\providecommand \enquote  [1]{``#1''}%
\providecommand \bibnamefont  [1]{#1}%
\providecommand \bibfnamefont [1]{#1}%
\providecommand \citenamefont [1]{#1}%
\providecommand \href@noop [0]{\@secondoftwo}%
\providecommand \href [0]{\begingroup \@sanitize@url \@href}%
\providecommand \@href[1]{\@@startlink{#1}\@@href}%
\providecommand \@@href[1]{\endgroup#1\@@endlink}%
\providecommand \@sanitize@url [0]{\catcode `\\12\catcode `\$12\catcode
  `\&12\catcode `\#12\catcode `\^12\catcode `\_12\catcode `\%12\relax}%
\providecommand \@@startlink[1]{}%
\providecommand \@@endlink[0]{}%
\providecommand \url  [0]{\begingroup\@sanitize@url \@url }%
\providecommand \@url [1]{\endgroup\@href {#1}{\urlprefix }}%
\providecommand \urlprefix  [0]{URL }%
\providecommand \Eprint [0]{\href }%
\providecommand \doibase [0]{https://doi.org/}%
\providecommand \selectlanguage [0]{\@gobble}%
\providecommand \bibinfo  [0]{\@secondoftwo}%
\providecommand \bibfield  [0]{\@secondoftwo}%
\providecommand \translation [1]{[#1]}%
\providecommand \BibitemOpen [0]{}%
\providecommand \bibitemStop [0]{}%
\providecommand \bibitemNoStop [0]{.\EOS\space}%
\providecommand \EOS [0]{\spacefactor3000\relax}%
\providecommand \BibitemShut  [1]{\csname bibitem#1\endcsname}%
\let\auto@bib@innerbib\@empty
\bibitem [{\citenamefont {{J.-G. Zhu and Chando Park}}(2006)}]{zhu_2006}%
  \BibitemOpen
  \bibfield  {author} {\bibinfo {author} {\bibnamefont {{J.-G. Zhu and Chando
  Park}}},\ }\bibfield  {title} {\bibinfo {title} {Magnetic tunnel junctions},\
  }\href@noop {} {\bibfield  {journal} {\bibinfo  {journal} {Mater. Today}\
  }\textbf {\bibinfo {volume} {9}},\ \bibinfo {pages} {36} (\bibinfo {year}
  {2006})}\BibitemShut {NoStop}%
\bibitem [{\citenamefont {{J. Heidmann and A. M.
  Taratorin}}(2011)}]{taratorin_2011_ch_1}%
  \BibitemOpen
  \bibfield  {author} {\bibinfo {author} {\bibnamefont {{J. Heidmann and A. M.
  Taratorin}}},\ }\bibfield  {title} {\bibinfo {title} {Magnetic recording
  heads},\ }in\ \href@noop {} {\emph {\bibinfo {booktitle} {Handbook of
  Magnetic Materials}}},\ Vol.~\bibinfo {volume} {19},\ \bibinfo {editor}
  {edited by\ \bibinfo {editor} {\bibfnamefont {K.~H.~J.}\ \bibnamefont
  {Buschow}}}\ (\bibinfo  {publisher} {Elsevier},\ \bibinfo {address}
  {Amsterdam, Netherlands},\ \bibinfo {year} {2011})\ \bibinfo {edition} {1st}\
  ed.,\ Chap.~\bibinfo {chapter} {1}, pp.\ \bibinfo {pages}
  {1--105}\BibitemShut {NoStop}%
\bibitem [{\citenamefont {Kittel}(1951)}]{kittel_1951}%
  \BibitemOpen
  \bibfield  {author} {\bibinfo {author} {\bibfnamefont {C.}~\bibnamefont
  {Kittel}},\ }\bibfield  {title} {\bibinfo {title} {Ferromagnetic resonance},\
  }\href@noop {} {\bibfield  {journal} {\bibinfo  {journal} {J. Phys. Radium.}\
  }\textbf {\bibinfo {volume} {12}},\ \bibinfo {pages} {291} (\bibinfo {year}
  {1951})}\BibitemShut {NoStop}%
\bibitem [{\citenamefont {Brataas}\ \emph {et~al.}(2012)\citenamefont
  {Brataas}, \citenamefont {Kent},\ and\ \citenamefont {Ohno}}]{brataas_2012}%
  \BibitemOpen
  \bibfield  {author} {\bibinfo {author} {\bibfnamefont {A.}~\bibnamefont
  {Brataas}}, \bibinfo {author} {\bibfnamefont {A.~D.}\ \bibnamefont {Kent}},
  and\ \bibinfo {author} {\bibfnamefont {H.}~\bibnamefont {Ohno}},\ }\bibfield
  {title} {\bibinfo {title} {Current-induced torques in magnetic materials},\
  }\href@noop {} {\bibfield  {journal} {\bibinfo  {journal} {Nat. Mater.}\
  }\textbf {\bibinfo {volume} {11}},\ \bibinfo {pages} {372} (\bibinfo {year}
  {2012})}\BibitemShut {NoStop}%
\bibitem [{\citenamefont {{D. V. Berkov and J. Miltat}}(2008)}]{berkov_2008}%
  \BibitemOpen
  \bibfield  {author} {\bibinfo {author} {\bibnamefont {{D. V. Berkov and J.
  Miltat}}},\ }\bibfield  {title} {\bibinfo {title} {Spin-torque driven
  magnetization dynamics: Micromagnetic modeling},\ }\href@noop {} {\bibfield
  {journal} {\bibinfo  {journal} {J. Magn. Magn. Mater.}\ }\textbf {\bibinfo
  {volume} {320}},\ \bibinfo {pages} {1238} (\bibinfo {year}
  {2008})}\BibitemShut {NoStop}%
\bibitem [{\citenamefont {{L. D. Landau and E. M.
  Lifshitz}}(1976)}]{landau_1976_ch_29}%
  \BibitemOpen
  \bibfield  {author} {\bibinfo {author} {\bibnamefont {{L. D. Landau and E. M.
  Lifshitz}}},\ }\bibinfo {title} {Course of theoretical physics: Mechanics}\
  (\bibinfo  {publisher} {Butterworth-Heinemann},\ \bibinfo {address} {Oxford,
  UK},\ \bibinfo {year} {1976})\ Chap.~\bibinfo {chapter} {29}, pp.\ \bibinfo
  {pages} {87--92},\ \bibinfo {edition} {3rd}\ ed.\BibitemShut {Stop}%
\bibitem [{\citenamefont {Maddaloni}\ \emph {et~al.}(2013)\citenamefont
  {Maddaloni}, \citenamefont {Bellini},\ and\ \citenamefont
  {Natale}}]{maddaloni_2013_ch_2_6_2}%
  \BibitemOpen
  \bibfield  {author} {\bibinfo {author} {\bibfnamefont {P.}~\bibnamefont
  {Maddaloni}}, \bibinfo {author} {\bibfnamefont {M.}~\bibnamefont {Bellini}},
  and\ \bibinfo {author} {\bibfnamefont {P.~D.}\ \bibnamefont {Natale}},\
  }\bibinfo {title} {Laser-based measurements for time and frequency domain
  applications: A handbook}\ (\bibinfo  {publisher} {CRC Press},\ \bibinfo
  {address} {Boca Raton, FL, USA},\ \bibinfo {year} {2013})\ Chap.\ \bibinfo
  {chapter} {2.6.2}, pp.\ \bibinfo {pages} {76--77}\BibitemShut {NoStop}%
\bibitem [{\citenamefont {Keatley}\ \emph {et~al.}(2016)\citenamefont
  {Keatley}, \citenamefont {Sani}, \citenamefont {Hrkac}, \citenamefont
  {Mohseni}, \citenamefont {D{\"u}rrenfeld}, \citenamefont {{\AA}kerman},\ and\
  \citenamefont {Hicken}}]{keatley_2016}%
  \BibitemOpen
  \bibfield  {author} {\bibinfo {author} {\bibfnamefont {P.~S.}\ \bibnamefont
  {Keatley}}, \bibinfo {author} {\bibfnamefont {S.~R.}\ \bibnamefont {Sani}},
  \bibinfo {author} {\bibfnamefont {G.}~\bibnamefont {Hrkac}}, \bibinfo
  {author} {\bibfnamefont {S.~M.}\ \bibnamefont {Mohseni}}, \bibinfo {author}
  {\bibfnamefont {P.}~\bibnamefont {D{\"u}rrenfeld}}, \bibinfo {author}
  {\bibfnamefont {J.}~\bibnamefont {{\AA}kerman}}, and\ \bibinfo {author}
  {\bibfnamefont {R.~J.}\ \bibnamefont {Hicken}},\ }\bibfield  {title}
  {\bibinfo {title} {Superharmonic injection locking of nanocontact spin-torque
  vortex oscillators},\ }\href@noop {} {\bibfield  {journal} {\bibinfo
  {journal} {Phys. Rev. B}\ }\textbf {\bibinfo {volume} {94}},\ \bibinfo
  {pages} {094404} (\bibinfo {year} {2016})}\BibitemShut {NoStop}%
\bibitem [{\citenamefont {Lebrun}\ \emph {et~al.}(2015)\citenamefont {Lebrun},
  \citenamefont {Jenkins}, \citenamefont {Dussaux}, \citenamefont {Locatelli},
  \citenamefont {Tsunegi}, \citenamefont {Grimaldi}, \citenamefont {Kubota},
  \citenamefont {Bortolotti}, \citenamefont {Yakushiji}, \citenamefont
  {Grollier}, \citenamefont {Fukushima}, \citenamefont {Yuasa},\ and\
  \citenamefont {Cros}}]{lebrun_2015}%
  \BibitemOpen
  \bibfield  {author} {\bibinfo {author} {\bibfnamefont {R.}~\bibnamefont
  {Lebrun}}, \bibinfo {author} {\bibfnamefont {A.}~\bibnamefont {Jenkins}},
  \bibinfo {author} {\bibfnamefont {A.}~\bibnamefont {Dussaux}}, \bibinfo
  {author} {\bibfnamefont {N.}~\bibnamefont {Locatelli}}, \bibinfo {author}
  {\bibfnamefont {S.}~\bibnamefont {Tsunegi}}, \bibinfo {author} {\bibfnamefont
  {E.}~\bibnamefont {Grimaldi}}, \bibinfo {author} {\bibfnamefont
  {H.}~\bibnamefont {Kubota}}, \bibinfo {author} {\bibfnamefont
  {P.}~\bibnamefont {Bortolotti}}, \bibinfo {author} {\bibfnamefont
  {K.}~\bibnamefont {Yakushiji}}, \bibinfo {author} {\bibfnamefont
  {J.}~\bibnamefont {Grollier}}, \bibinfo {author} {\bibfnamefont
  {A.}~\bibnamefont {Fukushima}}, \bibinfo {author} {\bibfnamefont
  {S.}~\bibnamefont {Yuasa}}, and\ \bibinfo {author} {\bibfnamefont
  {V.}~\bibnamefont {Cros}},\ }\bibfield  {title} {\bibinfo {title}
  {Understanding of phase noise squeezing under fractional synchronization of a
  nonlinear spin transfer vortex oscillator},\ }\href@noop {} {\bibfield
  {journal} {\bibinfo  {journal} {Phys. Rev. Lett.}\ }\textbf {\bibinfo
  {volume} {115}},\ \bibinfo {pages} {017201} (\bibinfo {year}
  {2015})}\BibitemShut {NoStop}%
\bibitem [{\citenamefont {Carpentieri}\ \emph {et~al.}(2013)\citenamefont
  {Carpentieri}, \citenamefont {Moriyama}, \citenamefont {Azzerboni},\ and\
  \citenamefont {Finocchio}}]{carpentieri_2013}%
  \BibitemOpen
  \bibfield  {author} {\bibinfo {author} {\bibfnamefont {M.}~\bibnamefont
  {Carpentieri}}, \bibinfo {author} {\bibfnamefont {T.}~\bibnamefont
  {Moriyama}}, \bibinfo {author} {\bibfnamefont {B.}~\bibnamefont {Azzerboni}},
  and\ \bibinfo {author} {\bibfnamefont {G.}~\bibnamefont {Finocchio}},\
  }\bibfield  {title} {\bibinfo {title} {Injection locking at zero field in two
  free layer spin-valves},\ }\href@noop {} {\bibfield  {journal} {\bibinfo
  {journal} {Appl. Phys. Lett.}\ }\textbf {\bibinfo {volume} {102}},\ \bibinfo
  {pages} {102413} (\bibinfo {year} {2013})}\BibitemShut {NoStop}%
\bibitem [{\citenamefont {Quinsat}\ \emph {et~al.}(2011)\citenamefont
  {Quinsat}, \citenamefont {Sierra}, \citenamefont {Firastrau}, \citenamefont
  {Tiberkevich}, \citenamefont {Slavin}, \citenamefont {Gusakova},
  \citenamefont {Buda-Prejbeanu}, \citenamefont {Zarudniev}, \citenamefont
  {Michel}, \citenamefont {Ebels}, \citenamefont {Dieny}, \citenamefont
  {Cyrille}, \citenamefont {Katine}, \citenamefont {Mauri}, ,\ and\
  \citenamefont {Zeltser}}]{quinsat_2011}%
  \BibitemOpen
  \bibfield  {author} {\bibinfo {author} {\bibfnamefont {M.}~\bibnamefont
  {Quinsat}}, \bibinfo {author} {\bibfnamefont {J.~F.}\ \bibnamefont {Sierra}},
  \bibinfo {author} {\bibfnamefont {I.}~\bibnamefont {Firastrau}}, \bibinfo
  {author} {\bibfnamefont {V.}~\bibnamefont {Tiberkevich}}, \bibinfo {author}
  {\bibfnamefont {A.}~\bibnamefont {Slavin}}, \bibinfo {author} {\bibfnamefont
  {D.}~\bibnamefont {Gusakova}}, \bibinfo {author} {\bibfnamefont {L.~D.}\
  \bibnamefont {Buda-Prejbeanu}}, \bibinfo {author} {\bibfnamefont
  {M.}~\bibnamefont {Zarudniev}}, \bibinfo {author} {\bibfnamefont {J.-P.}\
  \bibnamefont {Michel}}, \bibinfo {author} {\bibfnamefont {U.}~\bibnamefont
  {Ebels}}, \bibinfo {author} {\bibfnamefont {B.}~\bibnamefont {Dieny}},
  \bibinfo {author} {\bibfnamefont {M.-C.}\ \bibnamefont {Cyrille}}, \bibinfo
  {author} {\bibfnamefont {J.~A.}\ \bibnamefont {Katine}}, \bibinfo {author}
  {\bibfnamefont {D.}~\bibnamefont {Mauri}}, , and\ \bibinfo {author}
  {\bibfnamefont {A.}~\bibnamefont {Zeltser}},\ }\bibfield  {title} {\bibinfo
  {title} {Injection locking of tunnel junction oscillators to a microwave
  current},\ }\href@noop {} {\bibfield  {journal} {\bibinfo  {journal} {Appl.
  Phys. Lett.}\ }\textbf {\bibinfo {volume} {98}},\ \bibinfo {pages} {182503}
  (\bibinfo {year} {2011})}\BibitemShut {NoStop}%
\bibitem [{\citenamefont {Urazhdin}\ \emph {et~al.}(2010)\citenamefont
  {Urazhdin}, \citenamefont {Tabor}, \citenamefont {Tiberkevich},\ and\
  \citenamefont {Slavin}}]{urazhdin_2010}%
  \BibitemOpen
  \bibfield  {author} {\bibinfo {author} {\bibfnamefont {S.}~\bibnamefont
  {Urazhdin}}, \bibinfo {author} {\bibfnamefont {P.}~\bibnamefont {Tabor}},
  \bibinfo {author} {\bibfnamefont {V.}~\bibnamefont {Tiberkevich}}, and\
  \bibinfo {author} {\bibfnamefont {A.}~\bibnamefont {Slavin}},\ }\bibfield
  {title} {\bibinfo {title} {Fractional synchronization of spin-torque
  nano-oscillators},\ }\href@noop {} {\bibfield  {journal} {\bibinfo  {journal}
  {Phys. Rev. Lett.}\ }\textbf {\bibinfo {volume} {105}},\ \bibinfo {pages}
  {104101} (\bibinfo {year} {2010})}\BibitemShut {NoStop}%
\bibitem [{\citenamefont {Dussaux}\ \emph {et~al.}(2011)\citenamefont
  {Dussaux}, \citenamefont {Khvalkovskiy}, \citenamefont {Grollier},
  \citenamefont {Cros}, \citenamefont {Fukushima}, \citenamefont {Konoto},
  \citenamefont {Kubota}, \citenamefont {Yakushiji}, \citenamefont {Yuasa},
  \citenamefont {Ando},\ and\ \citenamefont {Fert}}]{dussaux_2011}%
  \BibitemOpen
  \bibfield  {author} {\bibinfo {author} {\bibfnamefont {A.}~\bibnamefont
  {Dussaux}}, \bibinfo {author} {\bibfnamefont {A.~V.}\ \bibnamefont
  {Khvalkovskiy}}, \bibinfo {author} {\bibfnamefont {J.}~\bibnamefont
  {Grollier}}, \bibinfo {author} {\bibfnamefont {V.}~\bibnamefont {Cros}},
  \bibinfo {author} {\bibfnamefont {A.}~\bibnamefont {Fukushima}}, \bibinfo
  {author} {\bibfnamefont {M.}~\bibnamefont {Konoto}}, \bibinfo {author}
  {\bibfnamefont {H.}~\bibnamefont {Kubota}}, \bibinfo {author} {\bibfnamefont
  {K.}~\bibnamefont {Yakushiji}}, \bibinfo {author} {\bibfnamefont
  {S.}~\bibnamefont {Yuasa}}, \bibinfo {author} {\bibfnamefont
  {K.}~\bibnamefont {Ando}}, and\ \bibinfo {author} {\bibfnamefont
  {A.}~\bibnamefont {Fert}},\ }\bibfield  {title} {\bibinfo {title} {Phase
  locking of vortex based spin transfer oscillators to a microwave current},\
  }\href@noop {} {\bibfield  {journal} {\bibinfo  {journal} {Appl. Phys.
  Lett.}\ }\textbf {\bibinfo {volume} {98}},\ \bibinfo {pages} {132506}
  (\bibinfo {year} {2011})}\BibitemShut {NoStop}%
\bibitem [{\citenamefont {Sankey}\ \emph {et~al.}(2006)\citenamefont {Sankey},
  \citenamefont {Braganca}, \citenamefont {Garcia}, \citenamefont {Krivorotov},
  \citenamefont {Buhrman},\ and\ \citenamefont {Ralph}}]{sankey_2006}%
  \BibitemOpen
  \bibfield  {author} {\bibinfo {author} {\bibfnamefont {J.~C.}\ \bibnamefont
  {Sankey}}, \bibinfo {author} {\bibfnamefont {P.~M.}\ \bibnamefont
  {Braganca}}, \bibinfo {author} {\bibfnamefont {A.~G.~F.}\ \bibnamefont
  {Garcia}}, \bibinfo {author} {\bibfnamefont {I.~N.}\ \bibnamefont
  {Krivorotov}}, \bibinfo {author} {\bibfnamefont {R.~A.}\ \bibnamefont
  {Buhrman}}, and\ \bibinfo {author} {\bibfnamefont {D.~C.}\ \bibnamefont
  {Ralph}},\ }\bibfield  {title} {\bibinfo {title} {Spin-transfer-driven
  ferromagnetic resonance of individual nanomagnets},\ }\href@noop {}
  {\bibfield  {journal} {\bibinfo  {journal} {Phys. Rev. Lett.}\ }\textbf
  {\bibinfo {volume} {96}},\ \bibinfo {pages} {227601} (\bibinfo {year}
  {2006})}\BibitemShut {NoStop}%
\bibitem [{\citenamefont {Tulapurkar}\ \emph {et~al.}(2005)\citenamefont
  {Tulapurkar}, \citenamefont {Suzuki}, \citenamefont {Fukushima},
  \citenamefont {Kubota}, \citenamefont {Maehara}, \citenamefont {Tsunekawa},
  \citenamefont {Djayaprawira}, \citenamefont {Watanabe},\ and\ \citenamefont
  {Yuasa}}]{tulapurkar_2005}%
  \BibitemOpen
  \bibfield  {author} {\bibinfo {author} {\bibfnamefont {A.~A.}\ \bibnamefont
  {Tulapurkar}}, \bibinfo {author} {\bibfnamefont {Y.}~\bibnamefont {Suzuki}},
  \bibinfo {author} {\bibfnamefont {A.}~\bibnamefont {Fukushima}}, \bibinfo
  {author} {\bibfnamefont {H.}~\bibnamefont {Kubota}}, \bibinfo {author}
  {\bibfnamefont {H.}~\bibnamefont {Maehara}}, \bibinfo {author} {\bibfnamefont
  {K.}~\bibnamefont {Tsunekawa}}, \bibinfo {author} {\bibfnamefont {D.~D.}\
  \bibnamefont {Djayaprawira}}, \bibinfo {author} {\bibfnamefont
  {N.}~\bibnamefont {Watanabe}}, and\ \bibinfo {author} {\bibfnamefont
  {S.}~\bibnamefont {Yuasa}},\ }\bibfield  {title} {\bibinfo {title}
  {Spin-torque diode effect in magnetic tunnel junctions},\ }\href@noop {}
  {\bibfield  {journal} {\bibinfo  {journal} {Nature}\ }\textbf {\bibinfo
  {volume} {438}},\ \bibinfo {pages} {339} (\bibinfo {year}
  {2005})}\BibitemShut {NoStop}%
\bibitem [{\citenamefont {{J.-G. Zhu and X. Zhu}}(2004)}]{zhu_2004}%
  \BibitemOpen
  \bibfield  {author} {\bibinfo {author} {\bibnamefont {{J.-G. Zhu and X.
  Zhu}}},\ }\bibfield  {title} {\bibinfo {title} {Spin transfer induced noise
  in {CPP} read heads},\ }\href@noop {} {\bibfield  {journal} {\bibinfo
  {journal} {IEEE Trans. Magn.}\ }\textbf {\bibinfo {volume} {40}},\ \bibinfo
  {pages} {182} (\bibinfo {year} {2004})}\BibitemShut {NoStop}%
\bibitem [{\citenamefont {{D. Berkov and N. Gorn}}(2005)}]{berkov_gorn_2005}%
  \BibitemOpen
  \bibfield  {author} {\bibinfo {author} {\bibnamefont {{D. Berkov and N.
  Gorn}}},\ }\bibfield  {title} {\bibinfo {title} {Transition from the
  macrospin to chaotic behavior by a spin-torque driven magnetization
  precession of a square nanoelement},\ }\href@noop {} {\bibfield  {journal}
  {\bibinfo  {journal} {Phys. Rev. B}\ }\textbf {\bibinfo {volume} {71}},\
  \bibinfo {pages} {052403} (\bibinfo {year} {2005})}\BibitemShut {NoStop}%
\bibitem [{\citenamefont {Roblin}(2011)}]{roblin_2011}%
  \BibitemOpen
  \bibfield  {author} {\bibinfo {author} {\bibfnamefont {P.}~\bibnamefont
  {Roblin}},\ }\href@noop {} {\emph {\bibinfo {title} {Nonlinear RF Circuits
  and Nonlinear Vector Network Analyzers}}},\ \bibinfo {edition} {1st}\ ed.\
  (\bibinfo  {publisher} {Cambridge University Press},\ \bibinfo {address}
  {Cambridge, UK},\ \bibinfo {year} {2011})\BibitemShut {NoStop}%
\bibitem [{\citenamefont {Remley}(2009)}]{remley_2009}%
  \BibitemOpen
  \bibfield  {author} {\bibinfo {author} {\bibfnamefont {K.~A.}\ \bibnamefont
  {Remley}},\ }\bibfield  {title} {\bibinfo {title} {Practical applications of
  nonlinear measurements},\ }in\ \href@noop {} {\emph {\bibinfo {booktitle}
  {Proc. 73rd ARFTG Microwave Measurement Conference}}}\ (\bibinfo  {publisher}
  {IEEE},\ \bibinfo {year} {2009})\ pp.\ \bibinfo {pages} {1--15}\BibitemShut
  {NoStop}%
\bibitem [{\citenamefont {Randall}(1987)}]{randall_1987_ch_2_2_1}%
  \BibitemOpen
  \bibfield  {author} {\bibinfo {author} {\bibfnamefont {R.~B.}\ \bibnamefont
  {Randall}},\ }\bibinfo {title} {Frequency analysis}\ (\bibinfo  {publisher}
  {Br{\"u}el \& Kj{\ae}r},\ \bibinfo {address} {N{\ae}rum, Denmark},\ \bibinfo
  {year} {1987})\ Chap.\ \bibinfo {chapter} {2.2.1}, p.~\bibinfo {pages} {23},\
  \bibinfo {edition} {3rd}\ ed.\BibitemShut {Stop}%
\bibitem [{\citenamefont {{R. E. Ziemer and W. H.
  Tranter}}(2014)}]{ziemer_2014_ch_2_6_4}%
  \BibitemOpen
  \bibfield  {author} {\bibinfo {author} {\bibnamefont {{R. E. Ziemer and W. H.
  Tranter}}},\ }\bibinfo {title} {Principles of communications: Systems,
  modulation, and noise}\ (\bibinfo  {publisher} {John Wiley \& Sons Ltd.},\
  \bibinfo {address} {USA},\ \bibinfo {year} {2014})\ Chap.\ \bibinfo {chapter}
  {2.6.4}, pp.\ \bibinfo {pages} {201--213},\ \bibinfo {edition} {7th}\
  ed.\BibitemShut {Stop}%
\bibitem [{\citenamefont {{D. V. Berkov and N. L. Gorn}}()}]{micromagus}%
  \BibitemOpen
  \bibfield  {author} {\bibinfo {author} {\bibnamefont {{D. V. Berkov and N. L.
  Gorn}}},\ }\href@noop {} {\bibinfo {title} {Micromagus: package for
  micromagnetic simulations}},\ \bibinfo {howpublished}
  {http://www.micromagus.de}\BibitemShut {NoStop}%
\bibitem [{\citenamefont {Slonczewski}(1996)}]{slonczewski_1996}%
  \BibitemOpen
  \bibfield  {author} {\bibinfo {author} {\bibfnamefont {J.~C.}\ \bibnamefont
  {Slonczewski}},\ }\bibfield  {title} {\bibinfo {title} {Current-driven
  excitation of magnetic multilayers},\ }\href@noop {} {\bibfield  {journal}
  {\bibinfo  {journal} {J. Magn. Magn. Mater.}\ }\textbf {\bibinfo {volume}
  {159}},\ \bibinfo {pages} {L1} (\bibinfo {year} {1996})}\BibitemShut
  {NoStop}%
\bibitem [{\citenamefont {Slonczewski}(2002)}]{slonczewski_2002}%
  \BibitemOpen
  \bibfield  {author} {\bibinfo {author} {\bibfnamefont {J.~C.}\ \bibnamefont
  {Slonczewski}},\ }\bibfield  {title} {\bibinfo {title} {Currents and torques
  in metallic magnetic multilayers},\ }\href@noop {} {\bibfield  {journal}
  {\bibinfo  {journal} {J. Magn. Magn. Mater.}\ }\textbf {\bibinfo {volume}
  {247}},\ \bibinfo {pages} {324} (\bibinfo {year} {2002})}\BibitemShut
  {NoStop}%
\bibitem [{\citenamefont {Xiao}\ \emph {et~al.}(2004)\citenamefont {Xiao},
  \citenamefont {Zangwill},\ and\ \citenamefont {Stiles}}]{xiao_2004}%
  \BibitemOpen
  \bibfield  {author} {\bibinfo {author} {\bibfnamefont {J.}~\bibnamefont
  {Xiao}}, \bibinfo {author} {\bibfnamefont {A.}~\bibnamefont {Zangwill}}, and\
  \bibinfo {author} {\bibfnamefont {M.~D.}\ \bibnamefont {Stiles}},\ }\bibfield
   {title} {\bibinfo {title} {Boltzmann test of {Slonczewski's} theory of
  spin-transfer torque},\ }\href@noop {} {\bibfield  {journal} {\bibinfo
  {journal} {Phys. Rev. B}\ }\textbf {\bibinfo {volume} {70}},\ \bibinfo
  {pages} {172405} (\bibinfo {year} {2004})}\BibitemShut {NoStop}%
\bibitem [{Note1()}]{Note1}%
  \BibitemOpen
  \bibinfo {note} {To describe the ST's angular dependence in MgO-based
  junctions, we used the common formulations for metallic junctions. Recent
  work (Ref.~\cite {kowalska_2018}) shows that the effective difference in
  angular dependence for these two scenarios is insignificant.}\BibitemShut
  {Stop}%
\bibitem [{\citenamefont {Jaffr{\`e}s}\ \emph {et~al.}(2001)\citenamefont
  {Jaffr{\`e}s}, \citenamefont {Lacour}, \citenamefont {{Van Dau}},
  \citenamefont {Briatico}, \citenamefont {Petroff},\ and\ \citenamefont
  {Vaur{\`e}s}}]{jaffres_2001}%
  \BibitemOpen
  \bibfield  {author} {\bibinfo {author} {\bibfnamefont {H.}~\bibnamefont
  {Jaffr{\`e}s}}, \bibinfo {author} {\bibfnamefont {D.}~\bibnamefont {Lacour}},
  \bibinfo {author} {\bibfnamefont {F.~N.}\ \bibnamefont {{Van Dau}}}, \bibinfo
  {author} {\bibfnamefont {J.}~\bibnamefont {Briatico}}, \bibinfo {author}
  {\bibfnamefont {F.}~\bibnamefont {Petroff}}, and\ \bibinfo {author}
  {\bibfnamefont {A.}~\bibnamefont {Vaur{\`e}s}},\ }\bibfield  {title}
  {\bibinfo {title} {Angular dependence of the tunnel magnetoresistance in
  transition-metal-based junctions},\ }\href@noop {} {\bibfield  {journal}
  {\bibinfo  {journal} {Phys. Rev. B}\ }\textbf {\bibinfo {volume} {64}},\
  \bibinfo {pages} {064427} (\bibinfo {year} {2001})}\BibitemShut {NoStop}%
\bibitem [{\citenamefont {Smith}\ \emph {et~al.}(2010)\citenamefont {Smith},
  \citenamefont {Carey},\ and\ \citenamefont {Childress}}]{smith_2010}%
  \BibitemOpen
  \bibfield  {author} {\bibinfo {author} {\bibfnamefont {N.}~\bibnamefont
  {Smith}}, \bibinfo {author} {\bibfnamefont {M.~J.}\ \bibnamefont {Carey}},
  and\ \bibinfo {author} {\bibfnamefont {J.~R.}\ \bibnamefont {Childress}},\
  }\bibfield  {title} {\bibinfo {title} {Measurement of {Gilbert} damping
  parameters in nanoscale {CPP-GMR} spin valves},\ }\href@noop {} {\bibfield
  {journal} {\bibinfo  {journal} {Phys. Rev. B}\ }\textbf {\bibinfo {volume}
  {81}},\ \bibinfo {pages} {184431} (\bibinfo {year} {2010})}\BibitemShut
  {NoStop}%
\bibitem [{\citenamefont {Pauselli}\ \emph {et~al.}(2017)\citenamefont
  {Pauselli}, \citenamefont {Stankiewicz}, \citenamefont {Zhang},\ and\
  \citenamefont {Carlotti}}]{pauselli_2017}%
  \BibitemOpen
  \bibfield  {author} {\bibinfo {author} {\bibfnamefont {M.}~\bibnamefont
  {Pauselli}}, \bibinfo {author} {\bibfnamefont {A.}~\bibnamefont
  {Stankiewicz}}, \bibinfo {author} {\bibfnamefont {Y.}~\bibnamefont {Zhang}},
  and\ \bibinfo {author} {\bibfnamefont {G.}~\bibnamefont {Carlotti}},\
  }\bibfield  {title} {\bibinfo {title} {Magnetic noise and spin-wave
  eigenmodes in a magnetic tunnel junction read head},\ }\href@noop {}
  {\bibfield  {journal} {\bibinfo  {journal} {IEEE Trans. Magn.}\ }\textbf
  {\bibinfo {volume} {53}},\ \bibinfo {pages} {4400606} (\bibinfo {year}
  {2017})}\BibitemShut {NoStop}%
\bibitem [{\citenamefont {Schroeder}(1970)}]{schroeder_1970}%
  \BibitemOpen
  \bibfield  {author} {\bibinfo {author} {\bibfnamefont {M.~R.}\ \bibnamefont
  {Schroeder}},\ }\bibfield  {title} {\bibinfo {title} {Synthesis of
  low-peak-factor signals and binary sequences with low autocorrelation},\
  }\href@noop {} {\bibfield  {journal} {\bibinfo  {journal} {IEEE Trans. Inf.
  Theory}\ }\textbf {\bibinfo {volume} {16}},\ \bibinfo {pages} {85} (\bibinfo
  {year} {1970})}\BibitemShut {NoStop}%
\bibitem [{\citenamefont {{Van der Ouderaa}}\ \emph {et~al.}(1988)\citenamefont
  {{Van der Ouderaa}}, \citenamefont {Schoukens},\ and\ \citenamefont
  {Renneboog}}]{van_der_ouderaa_1988}%
  \BibitemOpen
  \bibfield  {author} {\bibinfo {author} {\bibfnamefont {E.}~\bibnamefont {{Van
  der Ouderaa}}}, \bibinfo {author} {\bibfnamefont {J.}~\bibnamefont
  {Schoukens}}, and\ \bibinfo {author} {\bibfnamefont {J.}~\bibnamefont
  {Renneboog}},\ }\bibfield  {title} {\bibinfo {title} {Peak factor
  minimization of input and output signals of linear systems},\ }\href@noop {}
  {\bibfield  {journal} {\bibinfo  {journal} {IEEE Trans. Ins. Meas.}\ }\textbf
  {\bibinfo {volume} {37}},\ \bibinfo {pages} {201} (\bibinfo {year}
  {1988})}\BibitemShut {NoStop}%
\bibitem [{\citenamefont {{N. Smith and P. Arnett}}(2001)}]{smith_2001}%
  \BibitemOpen
  \bibfield  {author} {\bibinfo {author} {\bibnamefont {{N. Smith and P.
  Arnett}}},\ }\bibfield  {title} {\bibinfo {title} {White-noise magnetization
  fluctuations in magnetoresistive heads},\ }\href@noop {} {\bibfield
  {journal} {\bibinfo  {journal} {Appl. Phys. Lett.}\ }\textbf {\bibinfo
  {volume} {78}},\ \bibinfo {pages} {1448} (\bibinfo {year}
  {2001})}\BibitemShut {NoStop}%
\bibitem [{\citenamefont {Mohammadi}\ \emph {et~al.}(2017)\citenamefont
  {Mohammadi}, \citenamefont {Jones}, \citenamefont {Paul}, \citenamefont
  {Khodadadi}, \citenamefont {Mewes}, \citenamefont {Mewes},\ and\
  \citenamefont {Kaiser}}]{mohammadi_2017}%
  \BibitemOpen
  \bibfield  {author} {\bibinfo {author} {\bibfnamefont {J.~B.}\ \bibnamefont
  {Mohammadi}}, \bibinfo {author} {\bibfnamefont {J.~M.}\ \bibnamefont
  {Jones}}, \bibinfo {author} {\bibfnamefont {S.}~\bibnamefont {Paul}},
  \bibinfo {author} {\bibfnamefont {B.}~\bibnamefont {Khodadadi}}, \bibinfo
  {author} {\bibfnamefont {C.~K.~A.}\ \bibnamefont {Mewes}}, \bibinfo {author}
  {\bibfnamefont {T.}~\bibnamefont {Mewes}}, and\ \bibinfo {author}
  {\bibfnamefont {C.}~\bibnamefont {Kaiser}},\ }\bibfield  {title} {\bibinfo
  {title} {Broadband ferromagnetic resonance characterization of anisotropies
  and relaxation in exchange-biased {IrMn/CoFe} bilayers},\ }\href@noop {}
  {\bibfield  {journal} {\bibinfo  {journal} {Phys. Rev. B}\ }\textbf {\bibinfo
  {volume} {95}},\ \bibinfo {pages} {064414} (\bibinfo {year}
  {2017})}\BibitemShut {NoStop}%
\bibitem [{\citenamefont {Shivamoggi}(2014)}]{shivamoggi_2014_intro}%
  \BibitemOpen
  \bibfield  {author} {\bibinfo {author} {\bibfnamefont {B.~K.}\ \bibnamefont
  {Shivamoggi}},\ }\href@noop {} {\emph {\bibinfo {title} {Nonlinear Dynamics
  and Chaotic Phenomena: An Introduction}}},\ \bibinfo {edition} {2nd}\ ed.\
  (\bibinfo  {publisher} {Springer},\ \bibinfo {address} {Dordrecht,
  Netherlands},\ \bibinfo {year} {2014})\BibitemShut {NoStop}%
\bibitem [{\citenamefont {Slonczewski}(2005)}]{slonczewski_2005}%
  \BibitemOpen
  \bibfield  {author} {\bibinfo {author} {\bibfnamefont {J.~C.}\ \bibnamefont
  {Slonczewski}},\ }\bibfield  {title} {\bibinfo {title} {Currents, torques,
  and polarization factors in magnetic tunnel junctions},\ }\href@noop {}
  {\bibfield  {journal} {\bibinfo  {journal} {Phys. Rev. B}\ }\textbf {\bibinfo
  {volume} {71}},\ \bibinfo {pages} {024411} (\bibinfo {year}
  {2005})}\BibitemShut {NoStop}%
\bibitem [{\citenamefont {Krivorotov}\ \emph {et~al.}(2007)\citenamefont
  {Krivorotov}, \citenamefont {Berkov}, \citenamefont {Gorn}, \citenamefont
  {Emley}, \citenamefont {Sankey}, \citenamefont {Ralph},\ and\ \citenamefont
  {Buhrman}}]{krivorotov_2007}%
  \BibitemOpen
  \bibfield  {author} {\bibinfo {author} {\bibfnamefont {I.~N.}\ \bibnamefont
  {Krivorotov}}, \bibinfo {author} {\bibfnamefont {D.~V.}\ \bibnamefont
  {Berkov}}, \bibinfo {author} {\bibfnamefont {N.~L.}\ \bibnamefont {Gorn}},
  \bibinfo {author} {\bibfnamefont {N.~C.}\ \bibnamefont {Emley}}, \bibinfo
  {author} {\bibfnamefont {J.~C.}\ \bibnamefont {Sankey}}, \bibinfo {author}
  {\bibfnamefont {D.~C.}\ \bibnamefont {Ralph}}, and\ \bibinfo {author}
  {\bibfnamefont {R.~A.}\ \bibnamefont {Buhrman}},\ }\bibfield  {title}
  {\bibinfo {title} {Large-amplitude coherent spin waves excited by
  spin-polarized current in nanoscale spin valves},\ }\href@noop {} {\bibfield
  {journal} {\bibinfo  {journal} {Phys. Rev. B}\ }\textbf {\bibinfo {volume}
  {76}},\ \bibinfo {pages} {024418} (\bibinfo {year} {2007})}\BibitemShut
  {NoStop}%
\bibitem [{\citenamefont {Rippard}\ \emph {et~al.}(2010)\citenamefont
  {Rippard}, \citenamefont {Deac}, \citenamefont {Pufall}, \citenamefont
  {Shaw}, \citenamefont {Keller}, \citenamefont {Russek}, \citenamefont
  {Bauer},\ and\ \citenamefont {Serpico}}]{rippard_2010}%
  \BibitemOpen
  \bibfield  {author} {\bibinfo {author} {\bibfnamefont {W.~H.}\ \bibnamefont
  {Rippard}}, \bibinfo {author} {\bibfnamefont {A.~M.}\ \bibnamefont {Deac}},
  \bibinfo {author} {\bibfnamefont {M.~R.}\ \bibnamefont {Pufall}}, \bibinfo
  {author} {\bibfnamefont {J.~M.}\ \bibnamefont {Shaw}}, \bibinfo {author}
  {\bibfnamefont {M.~W.}\ \bibnamefont {Keller}}, \bibinfo {author}
  {\bibfnamefont {S.~E.}\ \bibnamefont {Russek}}, \bibinfo {author}
  {\bibfnamefont {G.~E.~W.}\ \bibnamefont {Bauer}}, and\ \bibinfo {author}
  {\bibfnamefont {C.}~\bibnamefont {Serpico}},\ }\bibfield  {title} {\bibinfo
  {title} {Spin-transfer dynamics in spin valves with out-of-plane magnetized
  {CoNi} free layers},\ }\href@noop {} {\bibfield  {journal} {\bibinfo
  {journal} {Phys. Rev. B}\ }\textbf {\bibinfo {volume} {81}},\ \bibinfo
  {pages} {014426} (\bibinfo {year} {2010})}\BibitemShut {NoStop}%
\bibitem [{\citenamefont {{D. V. Berkov and N. L. Gorn}}(2009)}]{berkov_2009}%
  \BibitemOpen
  \bibfield  {author} {\bibinfo {author} {\bibnamefont {{D. V. Berkov and N. L.
  Gorn}}},\ }\bibfield  {title} {\bibinfo {title} {Spin-torque driven
  magnetization dynamics in a nanocontact setup for low external fields:
  Numerical simulation study},\ }\href@noop {} {\bibfield  {journal} {\bibinfo
  {journal} {Phys. Rev. B}\ }\textbf {\bibinfo {volume} {80}},\ \bibinfo
  {pages} {064409} (\bibinfo {year} {2009})}\BibitemShut {NoStop}%
\bibitem [{\citenamefont {{D. V. Berkov and N. L. Gorn}}(2005)}]{berkov_2005}%
  \BibitemOpen
  \bibfield  {author} {\bibinfo {author} {\bibnamefont {{D. V. Berkov and N. L.
  Gorn}}},\ }\bibfield  {title} {\bibinfo {title} {Magnetization precession due
  to a spin-polarized current in a thin nanoelement: Numerical simulation
  study},\ }\href@noop {} {\bibfield  {journal} {\bibinfo  {journal} {Phys.
  Rev. B}\ }\textbf {\bibinfo {volume} {72}},\ \bibinfo {pages} {094401}
  (\bibinfo {year} {2005})}\BibitemShut {NoStop}%
\bibitem [{\citenamefont {Kowalska}\ \emph {et~al.}(2018)\citenamefont
  {Kowalska}, \citenamefont {Fukushima}, \citenamefont {Sluka}, \citenamefont
  {Fowley}, \citenamefont {K{\'a}kay}, \citenamefont {Aleksandrov},
  \citenamefont {Lindner}, \citenamefont {Fassbender}, \citenamefont {Yuasa},\
  and\ \citenamefont {Deac}}]{kowalska_2018}%
  \BibitemOpen
  \bibfield  {author} {\bibinfo {author} {\bibfnamefont {E.}~\bibnamefont
  {Kowalska}}, \bibinfo {author} {\bibfnamefont {A.}~\bibnamefont {Fukushima}},
  \bibinfo {author} {\bibfnamefont {V.}~\bibnamefont {Sluka}}, \bibinfo
  {author} {\bibfnamefont {C.}~\bibnamefont {Fowley}}, \bibinfo {author}
  {\bibfnamefont {A.}~\bibnamefont {K{\'a}kay}}, \bibinfo {author}
  {\bibfnamefont {Y.}~\bibnamefont {Aleksandrov}}, \bibinfo {author}
  {\bibfnamefont {J.}~\bibnamefont {Lindner}}, \bibinfo {author} {\bibfnamefont
  {J.}~\bibnamefont {Fassbender}}, \bibinfo {author} {\bibfnamefont
  {S.}~\bibnamefont {Yuasa}}, and\ \bibinfo {author} {\bibfnamefont {A.~M.}\
  \bibnamefont {Deac}},\ }\bibfield  {title} {\bibinfo {title} {Tunnel
  magnetoresistance angular and bias dependence enabling tuneable wireless
  communication},\ }\href@noop {} {\bibfield  {journal} {\bibinfo  {journal}
  {arXiv e-prints}\ ,\ \bibinfo {eid} {arXiv:1808.10812}} (\bibinfo {year}
  {2018})},\ \Eprint {https://arxiv.org/abs/1808.10812} {arXiv:1808.10812
  [physics.app-ph]} \BibitemShut {NoStop}%
\end{thebibliography}%

\end{document}